\documentclass[acmsmall]{acmart}

\AtBeginDocument{%
  \providecommand\BibTeX{{%
    \normalfont B\kern-0.5em{\scshape i\kern-0.25em b}\kern-0.8em\TeX}}}

\setcopyright{rightsretained}
\copyrightyear{2024}
\acmYear{2024}
\acmDOI{10.1145/1122445.1122456}

\usepackage{colortbl} 

\usepackage{todonotes}
\usepackage{makecell}

\def\myBullet{\color[HTML]{656565}$\bullet$}

\begin{document}

\title{How Could AI Support Design Education?\newline A Study Across Fields Fuels \textit{Situating Analytics}}

\author{Ajit Jain}
\authornote{Current affiliation: Audigent}
\email{ajain24@tamu.edu}
\author{Andruid Kerne}
\email{andruid@ecologylab.net}
\authornote{Current affiliation: University of Illinois Chicago}
\author{Hannah Fowler}
\authornote{Current affiliation: Microsoft}
\author{Jinsil Seo}
\author{Galen Newman}
\author{Nic Lupfer}
\authornote{Current affiliation: Mapware}
\author{Aaron Perrine}
\authornote{Current affiliation: Independent}
\affiliation{%
  \institution{Texas A\&M University}
  \country{USA}
}

\renewcommand{\shortauthors}{Jain et al.}

\begin{abstract}
We use the process and findings from a case study of design educators' practices of assessment and feedback to fuel theorizing about how to make AI useful in service of human experience. We build on Suchman's theory of situated actions. We perform a qualitative study of 11 educators in 5 fields, who teach design processes situated in project-based learning contexts. Our study methodology evolves to incorporate aspects of co-design. Drawing on Charmaz's approach to grounded theory qualitative data gathering and analysis, we derive codes: design process; assessment and feedback challenges; and computational support. We focus our presentation of findings and derivation of implications for design around the research question: how could AI support assessment and feedback in design education?

We twice invoke creative cognition’s \textit{family resemblance principle}. First, family resemblance explains how design instructors already use assessment rubrics: no particular trait is necessary or sufficient; each only tends to indicate good design work. Second, family resemblance explains the analogous role for design creativity analytics: to measure characteristics that tend to indicate good design work, again, without being necessary or sufficient. Human teachers remain essential. We develop a set of situated design creativity analytics\textemdash Fluency, Flexibility, Visual Consistency, Multiscale Organization, and Legible Contrast\textemdash in order to support instructors' efforts, by providing on-demand, learning objectives-based assessment and feedback to students.

We theorize a methodology, which we call \textit{situating analytics}, firstly because making AI support living human activity depends on aligning what analytics measure with situated practices. Further, we realize that analytics can become most significant to users by situating them through interfaces that integrate them into the material contexts of their use. Here, this means situating design creativity analytics into actual design environments, in order to make visible relationships between particular design elements and their assemblages with specific analytics that describe and measure them. Through the case study, we identify situating analytics as a methodology for explaining analytics to users, because the iterative process of alignment with practice has the potential to enable data scientists to derive analytics that make sense as part of and support situated human experiences.
\end{abstract}



\maketitle

\section{Introduction}
How could it be possible for AI to support human design instructors in assessing and providing feedback to students studying design? 
As students learn to develop creative solutions for design problems, they need frequent assessment and feedback to make progress \cite{oh2013theoretical}.
Design problems are known as \emph{wicked}, that is, addressing the people and needs at hand; replete with confusing information and conflicting stakeholders' values; open to multiple explanations; and formulated and reformulated not absolutely, but in terms of a designer's conception \cite{rittel1973dilemmas,cross1982designerly,buchanan1992wicked,zimmerman2007research, stolterman2008nature}. 
Design students are appropriately challenged by this wickedness. 
They sometimes struggle to the extent that they drop out \cite{osmond2015threshold}.

To help students learn in the face of wicked problems, design instructors structure courses to provide frequent, helpful feedback \cite{dannels2011students}.
However, as design education demands continue to grow, instructors face challenges in providing timely assessment and feedback \cite{krause2017critique}.
Further, instructors may not be available at critical times, e.g., late the night before an assignment is due. 
These conditions translate to needs for alternative channels of assessment and feedback.
Human assistance channels include peer \cite{papadopoulos2014community,mcdonald2019perceived} and crowdsourced \cite{yuan2016almost,dow2013pilot,hui2014crowd} design feedback.
The present research alternatively focuses on the potential of new forms of computation, which complement the human assistance by providing the ability to process big data at speed \cite{kittur2019scaling}. 

To supplement instructor efforts, under time pressure, our objective is to derive implications for AI-based design analytics and their presentation via dashboards.
Like recent AI-based analytics research \cite{how2020artificial,lim2018design}, we adopt a broad view of analytics, which includes not only facts but also inferences.
While learning analytics have been found useful in various courses, prior work lacks in investigating their efficacy in design education contexts.
This does not mean that computational modeling of creative design, which can constitute a basis for AI and analytics, is impossible.
Gero and Maher developed a computational model for supporting creative design based on the application of analogy and mutation processes to design representations \cite{gero1991mutation}.
Reinecke et al. demonstrated the potential of computation to assess aesthetic quality of website design \cite{reinecke2013predicting}. 
Oulasvirta et al. provide a web service that assesses a graphical user interface design against a variety of metrics, ranging from symmetry to colorfulness to visual clutter \cite{oulasvirta2018aalto}.
This article addresses the research gap of how to build AI support for design courses.

We argue that AI support for design education must provide transparent, on demand assessment and feedback.
How can this work?
Suchman seminally articulated that designing a useful AI system requires understanding users' \textit{situated practice}, i.e., how their actions develop purpose and intelligibility within particular circumstances.
Suchman emphasizes a system's transparency\textemdash to convey AI's intended purpose to the users and establish its accountability\textemdash as requisite for effectively supporting situated practice \cite{suchman1987plans}. 
Dourish breaks this principle down to focus on how translating ideas between intellectually different domains of situated practice (social) and technology (computational) ``can be both exceptionally valuable and unexpectedly difficult'' \cite{dourish2004we}. 
While focused on user experience and HCI, we find that recent studies by Yang et al. \cite{yang2020re} and Dove et al. \cite{dove2017ux} corroborate the difficulty of translating ideas between the domains of \textit{design} practice and \textit{AI} technology. 
Dourish argues that making contextual properties transparent is vital in addressing the difficulty, and as context acquires ``meaning or relevance through their relationship to forms of practice'', study of work practices becomes fundamentally important \cite{dourish2004we}. 

The present research develops a case study for human-centered AI, using qualitative, co-design methods to undertake the `difficult' work of understanding practices in design education, in order to derive a basis of `contextual properties' for AI support.
While prior work has studied assessment contexts in design education, and identified a range of criteria \cite{de2009assessment}, it has not detailed \textit{how} instructors apply these criteria in practice. 
Hence, to derive implications for computational support, we investigated how design instructors, across fields\textemdash e.g., architecture, interactive art \& design, mechanical engineering, and computer science\textemdash perform assessment.
In our study contexts, these courses are taught by each field’s faculty, in contrast with dedicated design programs\textemdash e.g., TU Eindhoven, TU Delft, and Stanford d.school\textemdash which teach students from multiple disciplines \cite{wrigley2017design}.
The present investigation began with the goal of addressing this research question:

\textbf{RQ1}: How, if at all, could analytics and AI support instructors in assessing and providing feedback on design?

Early in our investigation, we realized that co-design methods are imperative, to elicit and build on tacit and contextualized knowledge and practices of stakeholders \cite{sanders2012convivial,sanders2008co}.
Our focus evolved, by performing the research, through co-design engagements. 
Working closely with design instructors, we discovered another, emergent and underlying research question: 

\textbf{RQ2}: What challenges characterize how instructors teach, assess, and provide feedback on design (across fields)?

We present prior work relevant to various aspects of our investigation.
Then, we describe our study methodology, including aspects of co-design for involving instructors, and a grounded theory approach for analyzing qualitative data acquired through co-design engagements.
We follow this by presenting findings and discussing challenges, which characterize assessment and feedback across the diverse design courses of our study.
The particulars of these situated understandings and needs for using AI to support design education lead us to next contribute a generalized methodology for situating AI-based analytics. 
We invoke this methodology to derive potential new forms of AI support for design education.
We reflect on the value of and role for co-design methods for situating AI support.
We conclude by reiterating theoretical contributions for human-centered AI, in general, and design education, as a particular situated domain.

\section{Prior Work}
We begin by presenting creative cognition research, building on which researchers have developed effective approaches for measuring design creativity.
We note how the present research draws on creative cognition's family resemblance principle.
We follow with prior work investigating design education assessment and feedback practices and identify gaps in current understandings of instructors' approaches.
We then note the dearth of learning technologies for design education and distinguish the present research from prior computational support investigations for assessing and providing feedback on design.

\subsection{Creative Cognition: Family Resemblance Principle}
\label{sec:sensitizing_creative_cognition}

Ideation is the creative process of developing and generating ideas. 
Design is a general activity oriented toward creative ideation \cite{frich2018twenty}.
Hence, understanding how to measure creativity can prove beneficial in developing approaches for assessing students' design work.
At the same time, as creativity researchers discuss, “the exact question of what is creativity is often ignored or answered in too many different ways” \cite{kaufman2009beyond}.  
Frich et al.’s recent survey recommends “looking to the well-established tradition of psychology-based creativity research” \cite{frich2018twenty}.

In cognitive psychology, creative cognition is a field of investigating processes and structures that contribute to creative thinking. 
Some creative processes are exploratory\textemdash e.g., attribute finding, conceptual interpretation, and functional inference\textemdash while others are generative\textemdash e.g., memory retrieval, association, and mental synthesis \cite{smith1995creative}.  
Creative structures are representations, such as novel visual patterns, verbal combinations, and mental blends. 
Building on creative cognition theory, researchers have derived and applied ideation metrics for measuring design creativity in particular contexts, including Fluency, Flexibility/Variety, Novelty, Emergence and Visual Presentation \cite{shah2003metrics,kerne2014using}.

Creative cognition avoids the epistemological trap of defining creativity in any absolute way. 
Alternatively, it presents the \textit{family resemblance} principle, which states: while cognitive processes and structures are indicators, \textit{no particular process or structure is necessary or sufficient for creative ideation}. 
The present research applies this principle to develop new understanding of how design instructors use assessment rubrics and theorize the validity of conceptually aligned AI-based analytics.
It then develops novel, AI-based approaches for measuring design creativity.

\subsection{Design Education: Assessment and Feedback Practices}
Design education, across fields, is based on common pedagogical beliefs and practices \cite{sawyer2018teaching,jain2021support}.
Instructors define learning outcomes, project assignments, and rubrics, but keep them open-ended and flexible for students to explore and develop their own creative solutions \cite{sawyer2017teaching}.
Instructors engage students in prototyping \cite{brandt2013theoretical}, divergent thinking \cite{crismond2012informed,sawyer2018teaching}, extensive critique \cite{dannels2008beyond,oh2013theoretical}, and working closely with end-users \cite{crismond2012informed,vyas2013creative}. 
Design pedagogy inducts students into a \textit{community of practice} \cite{lave1991situated}, where they learn and develop skills, often through socially situated understandings of work that practitioners perform \cite{shreeve2010kind}.

Design instructors assess student work against a range of \textit{process}\textemdash e.g., problem solving, analytical thinking, abstract thinking, synthesis, creative thinking, and decision making\textemdash and \textit{product}\textemdash e.g., concept, layout, color, shape, texture, and rhythm\textemdash characteristics \cite{de2009assessment}.
Students need to develop both group\textemdash e.g., teamwork and communication\textemdash and individual\textemdash e.g., abilities to act independently and meta-cognitive skills to assess their own actions\textemdash competences \cite{gray2017individual}.
While this prior work enumerates a range of characteristics, it does not surface instructors' approaches to assessing these characteristics.
Without this contextual knowledge, as per Suchman and Dourish (above), it is not possible to derive implications for analytics and AI support.
Hence, the present research focuses on addressing this gap.

We next draw on prior work addressing how instructors provide structure to elicit and facilitate students' learning and expertise through feedback.
Feedback based on assessment of various design characteristics plays an important role, for advancing design students’ development of expertise \cite{oh2013theoretical}. 
As students begin, they may be unfamiliar with the open-endedness of the design work \cite{chen2016exploring}.
Many students have trouble working \cite{lande2010difficulties} on wicked design problems \cite{buchanan1992wicked} that require ongoing formulation, while dealing with confusing information and stakeholders’ conflicting values.
Instructors scaffold students’ mastery of creative processes and development of key competencies \cite{mcdonnell2016scaffolding}.
They use a “suggest, don’t tell” approach in their feedback, thus simultaneously directing students’ attention to key aspects and empowering them to act independently \cite{adams2016characterizing}.

To facilitate feedback, design courses schedule critique sessions, where instructors and peers help students reflect on their work \cite{oh2013theoretical,dannels2011students,kolko2012transformative}.
Critiques usually take place in studio settings: ``active sites where students are engaged intellectually and socially, shifting between analytic, synthetic, and evaluative modes of thinking in different sets of activities'' \cite{dutton1987design}.
Critiques range from frequent, informal “desk crits” and “pin ups” to more formal “jury” reviews at the end of a project \cite{oh2013theoretical,dannels2011students}.
They provide useful feedback to students and help them develop communication competencies—e.g., demonstration of design evolution, interaction with audience, and credible staging of presentation performance—through understandings of disciplinary norms, expectations, and behaviors \cite{dannels2008beyond}.  
In addition to public critique, instructors and peers engage through document collaborations and backchannel chat \cite{gray2018democratizing}. 

As instructor and peer availability are limited, additional feedback channels have been investigated.
For example, prior work focused on the potential of crowdsourcing \cite{yuan2016almost,dow2013pilot,luther2015structuring,hui2014crowd}.
We alternatively focus on computational approaches, which have received little attention in this regard.

\subsection{Computational Support: Design Assessment and Feedback}
As we argue in the introduction, developing computational support that provides transparent assessment and feedback in design course contexts requires understanding situated practice. 
Prior studies investigated practices across design courses, presenting findings on critique \cite{dannels2008beyond,oh2013theoretical,mcdonald2019perceived}, criteria \cite{de2009assessment}, and focused on process, rather than products \cite{greene2019fostering}. 
Implications for computational support were not addressed.
We observe that, because design creativity is so based in contextualized human experience, developing computational support for design and its assessment seems almost inimical. 
We are inspired by other researchers, who studied design work practices and presented meaningful implications for computational support, such as using the internet of things (IoT) to assist with interaction and communication processes \cite{vyas2013creative}.

Specifically, in relation to research question RQ1, we note that learning technologies, such as analytics and dashboards, have proven useful in supporting instructor assessment and feedback \cite{dawson2012using,arnold2012course,verbert2014learning}.
In lecture-based contexts, learning analytics\textemdash e.g., the number of times a student accessed a resource, time spent, and length of textual annotations\textemdash have assisted instructors in evaluating student understanding \cite{dawson2012using}.
For design course contexts, which involve project-based learning, there is a dearth of useful insights.
The present research identifies the underlying challenges and develops ideas for building analytics and dashboards suited for project-based design education.

As we briefly state in the introduction, researchers have developed highly accurate computational models for design assessment, which constitute a basis for deriving AI-based analytics corresponding to instructors' rubrics.
For example, Reinecke et al. assessed website aesthetics by developing a model based on features such as colors, number of words, and number of images \cite{reinecke2013predicting}.
Mackeprang et al. assess diversity by computing semantic similarity among ideas, using external knowledge graphs such as Wikidata and DBpedia \cite{mackeprang2018concept}.
Oulasvirta et al. assess a GUI design against a range of characteristics, such as the number of unique colors, color clustering, edge density, and whitespace  \cite{oulasvirta2018aalto}.
However, only a limited number of systems make assessment transparent to the user.
For example, Oulasvirta et al. inform the user the rationale behind the assessed characteristics, by including references to theoretical bases for ideas, and in some cases, visualizing the characteristic.
However, these systems have not been investigated in design course contexts.

To address how AI could support design education, the present research meets the gap at the intersection of two bodies of research: one that studies situated practice in design course contexts, and other that develops computational support for assessment of design characteristics.

Computational support for providing feedback on design has been investigated as well. 
Krause et al. crowdsourced labeled examples of feedback on student designs, and then used a natural language processing model to provide suggestions and improve crowdsourcing feedback on new designs \cite{krause2017critique}.
Ngoon et al. developed a text classifier to categorize feedback as actionable, specific, and/or justified.
They used the categorization to suggest examples to the reviewers, which improved the quality of feedback \cite{ngoon2018interactive}. 
Unlike assessment systems mentioned further above, both these feedback systems were investigated in design course contexts.
The present research makes complementary contributions by developing ideas for feedback sourced from AI assessment, as an alternative to the crowd, as well as by identifying and addressing challenges in tracking the incorporation of feedback.

\section{Methodology}
\label{sec:methodology}
\textit{\textbf{We}} engaged aspects of co-design \cite{sanders2012convivial,sanders2008co} to develop understandings of assessment and feedback practices and needs in a range of design course contexts, across fields. 
``We'' here refers to the resulting research team, comprised of the “initial research team” (IRT) members plus 2 design instructors who later chose to get more involved in the research, through the co-design process.
The IRT is a group of HCI researchers, consisting of the PI, and graduate and undergraduate students. 
The PI additionally participated as a design instructor. 
The remainder of the section differentiates between this larger we, and the IRT, in describing processes and methods.

The IRT employed a co-design approach, with the goal of building on tacit, shared knowledge and communication involving users, as participants, so that outcomes support situated use.
To initiate the co-design process, the IRT recruited other design instructors via email, starting with whose work they were aware of from past interdisciplinary collaborations. 
Based on instructors’ recommendations, the IRT performed snowball sampling \cite{biernacki1981snowball}. 
They motivated new participants by identifying explicit, immediate research goals of value to them, such as opportunities to reflect on their own processes and share perspectives with peers. 
IRT engaged instructors in dialogue about course learning objectives, assignment specifications, and assessment methods (See appendix \ref{sec:discussion_topics}).
Furthermore, they discussed whether and how computational means could support them, in particular, in design assessment.

The IRT initiated a \textit{grounded theory} approach\textemdash based on the work of Charmaz \cite{charmaz2014constructing}\textemdash to data collection and analysis. 
In this approach, grounded theory is a rigorous qualitative method, wherein researchers perform constant comparisons among pieces of data, to develop analytical codes and categories.
The method begins with sensitizing concepts\textemdash i.e., background ideas that provide researchers framing and guidance\textemdash and (cross) disciplinary perspectives, which inform the formulation of opening research questions. 
These, in turn, inform data collection and, the initial qualitative coding (labeling) of data elements.
Initial coding is followed by focused coding, in which codes become merged and rearticulated.
Conceptual categories get refined.
Directions emerge through this continuous, comparative qualitative analysis.
The sensitizing concepts and research questions continue to feed back, as researchers inductively interpret the codes and categories to derive theory that explains investigated phenomena \cite{charmaz2014constructing}.

{
\renewcommand{\arraystretch}{1.3} 
\begin{table*}[t!]
\center
\footnotesize
\begin{tabular}{c|c|c|c|c}
{\bf ID} & {\bf Gender} & {\bf Field} & {\bf Title} & {\bf UG Course Level} \\
\hline
D1  & F & Architecture & Associate Professor & Freshman, Sophomore, Senior \\ \rowcolor[gray]{0.95}
D2  & F & Interactive Art \& Design & Associate Professor & Junior, Senior \\
D3  & M & Architecture & Professor & Sophomore, Senior \\ \rowcolor[gray]{0.95}
D4  & F & Interactive Art \& Design & Assistant Professor & Junior, Senior \\
D5  & M & \makecell{Landscape Architecture \\ \& Urban Planning} & Assistant Professor & Sophomore, Junior \\ \rowcolor[gray]{0.95}
D6  & M & \makecell{Landscape Architecture \\ \& Urban Planning} & Associate Professor & Sophomore, Junior \\
D7  & M & Architecture & Dean & Freshman \\ \rowcolor[gray]{0.95}
D8  & M & Mechanical Engineering & Professor & Senior \\
D9  & F & Mechanical Engineering & Lecturer & Sophomore, Senior \\ \rowcolor[gray]{0.95}
D10 & F & Mechanical Engineering & Assistant Professor & Senior \\
D11 & M & \makecell{Computer Science \\ and Engineering} & Professor & Junior \\
\end{tabular}
\caption{Our study participants consist of design instructors from diverse fields. 
We discussed course learning objectives, assignment specification, and assessment and feedback methods.
With the exception of D7, D9, and D10, all teach graduate level courses, in addition to the undergraduate levels included above.
}
\label{tab:participants}
\end{table*}
}

Our discussions resulted in ongoing discourse, spread through multiple sessions, over a period of a year, as our understanding of teaching and assessment practices evolved by talking to different instructors. 
Emergent discussion topics include students’ use of sketching, instructors’ creativity assessment, and students’ considering instructor feedback beyond a specific deliverable. 
We refer to these sessions as discussions, because the IRT shared perspectives gained from prior work and by talking with other instructors. In total, 11 instructors from 2 universities\textemdash from 5 fields\textemdash participated in these serial discussions (Table \ref{tab:participants}). 
During these discussions, the IRT asked instructors for example design assignments, including feedback provided to students, developing context.

Before proceeding further, we note that, in our research, like others \cite{sjoberg1998participatory,luck2003dialogue,frauenberger2011designing}, co-design methods and grounded theory are complementary approaches.
Co-design supports designers and stakeholders not necessarily trained in design to work together in defining research products and designing solutions.
This involvement generates data.
However, co-design does not prescribe a particular data analysis approach.
For data analysis, we use grounded theory methods, which enable discoveries, both expected and unexpected, to emerge through data.

The IRT organized a Design Assessment Workshop, bringing together 5 instructors, for an interdisciplinary discussion of situated practice. These included 2 instructors from landscape architecture, 1 from interactive art \& design, 1 from mechanical engineering, and 1 from computer science. The number of participants was limited by scheduling constraints.

For the workshop, the IRT asked each participating instructor to present one of their design ideation deliverables, along with their methods of assessment. 
Participants engaged in extensive discussion, sharing perspectives based on practices in their field and classroom experiences. 
Together, participants identified common pain points and potential solutions. 
Participants responded with periodic sighing and murmurs of enthusiasm. 
Topics included the sizes of courses, teaching teams, and student teams; the level of details in assignment and rubric specifications; challenges in tracking students’ progress and incorporation of feedback; and, difficulty assessing contributions in team assignments. The workshop thus fostered developing comparative understanding and co-creation of shared meanings, goals, and outcomes. 
Finding value in the research, two instructors chose to participate further, by writing this paper as co-authors. 
We further discuss how interests, understandings, stakes, and collaboration evolved in the section \textit{Co-Design: Build a Design Education Stakeholders Community} (Section \ref{sec:pd_building_community}).

Two members of the IRT transcribed the individual and workshop discussion recordings and then performed qualitative codings of the transcribed data. 
Using Charmaz’ approach \cite{charmaz2014constructing}, first, they performed initial coding of 3 individual discussions. 
Then, the whole IRT met to make the codes consistent and bring them into alignment. Next, we performed focused coding of remaining data. Based on relationships among codes, we organized them into categories. In a subsequent step, the larger we, including the two participating instructor co-authors, revised the codes, as needed, to suitably represent salient themes.

We next present findings, integrated with discussion, and implications for new forms of computation to address needs and provide support. 
We derived 3 categories through our grounded theory analysis of qualitative data, guided by our research questions and sensitizing concepts: 1) Design Process, 2) Assessment and Feedback Challenges, and 3) Computational Support. 
As this article addresses how AI could support design education, which hinges on assessment and feedback across fields, here we only present themes from the second category and those pertaining to assessment and feedback in the third.
We detail only those findings that complement prior work.
We feature participant quotes illustrating phenomena.

\section{Assessment and Feedback Challenges: Findings + Discussion}
We present and interpret findings involving how instructors use rubrics of criteria in the assessment of student design work.
Instructors specify grading rubrics, in accordance with learning objectives, through which students demonstrate performance \cite{davis2017teaching}.  
Instructors’ rubrics for design assessment assign weights to manifestations of student work processes, as well as resulting products.
We contribute applying creative cognition’s family resemblance principle to understand how rubrics work in practices, their functions and limitations. 
We problematize issues that arise in assessing contributions to team projects.

Feedback based on the assessment of various design characteristics plays a vital role in helping students make progress. 
We discuss various forms of feedback, e.g., verbal and redlining.
We discuss challenges that arise both for instructors, who need to frequently provide feedback, and for students, who need to incorporate it. 

\subsection{Rubrics of Criteria and their Limits}
Assessment criteria, in the form of grading rubrics, operationalize a variety of design characteristics that instructors find important for students to learn and demonstrate in an assignment. 
As D8 puts, criteria are \textit{“kinds of rules that you can turn into a rubric.”} 
In comparison with fact-oriented \cite{erickson2002concept} assessment, such as through examinations that involve memorizing and reproducing, design students are required to demonstrate ability to reflect, apply, and understand \cite{tang1999students}.  

D2: \textit{I look at not just how well-functioning something is...conceptually, it meets what you planned...and also technically, it’s been implemented correctly and functioning right. Also, aesthetics.}

When assessing, instructors in our study are interested not only in design product outcomes, but further, in the processes students perform to achieve them. 
Like de la Harpe et al. \cite{de2009assessment}, we found that the weights, which product and process are assigned, vary across contexts. 
In line with Greene et al.’s work \cite{greene2019fostering}, we observe the need for focus on the process.

D8: \textit{We tend to grade them on the process and if the process is coherent, but we also try to look at the outcome...In a professional context, you can look at products...and say this is a good design or not. But in a pedagogical context, maybe we shouldn’t do this.}

We found that instructors in different fields, in this study, focus on different design characteristics, e.g., landscape architecture emphasizes visual aspects, mechanical engineering is more functionally oriented, and art \& design emphasizes interaction aspects.
D11’s computer science and engineering assignments simultaneously focus on visual, interaction, and functional design characteristics.
These differences in emphasis, on diverse design characteristics, is consistent both with prior work \cite{yilmaz2016feedback} and with instructors statements: design lacks universal criteria.
Assessing it can be subjective, because in the end, some level of personal preference gets involved. 

\emph{Consistently assessing design projects is thus, in itself a wicked problem.}
There are many variables to consider. The projects differ in objectives, personalities within teams, tastes of instructors, and situated aspects of sponsor capability and communication. Even when teams work on the same problem, their approaches may significantly differ. In D8’s words, \textit{“There’s no right answer.”} Instructors create rubrics to provide structure. At the same time, recognizing and assessing the uniqueness of projects is vital. Project deliverables, criteria, and rubrics become amenable to change \cite{sawyer2017teaching}. 

We contribute a new understanding that creative cognition’s \textit{family resemblance principle} \cite{smith1995creative} models the way that rubrics function in design courses. Instructors specify rubrics, but don’t generally believe that particular elements of their rubrics are exactly necessary or sufficient. Rather, good solutions to the design problems tend to exhibit characteristics that the rubrics specify.

D5: \textit{This is one of those hard things, because...some students could just do something really simple and it’s just like, “Uh yes, that’s it”. And somebody else could have something really complex and you’re like, “Yes, that’s it”.}

We use this new understanding, of how rubrics work according to the family resemblance principle, in conjunction with ideas from instructors, to inform design of computational support, such as AI-based design analytics and dashboards (Section \ref{sec:implications}).

\subsection{Assessing Contributions to Team Projects}
We find that assessing contributions in team projects is challenging and problematic. 
Sometimes some students get ‘floated’ in a course: other team members’ efforts carry them. 
It becomes difficult for instructors to assess contributions.

The topic became a major discussion point during the workshop.
One of the participating instructors brought this idea up, while envisioning computational means for providing insights into students' design processes.
Soon, every instructor present in the workshop started identifying with the pain point and shared similar experiences.

D2: \textit{All the roles are defined by the team. I kind of go and check who’s doing what, but still, if the team is all agreed on certain each individual’s role, they’re happy with it, I can’t really punish the individual person.}

D9: \textit{I had 36 teams this semester and saw different kinds of team dynamic issues...I would be interested in things that help us in maybe more objectively gauge student contributions to the project.}

On the one hand, these problems point to social loafing where members make less efforts as responsibility is diffused and social matching where members conform to ideas \cite{kreijns2003identifying,kohn2011collaborative}.
Creative cognition studies have shown that this reduces diversity of ideas \cite{kohn2011collaborative}. 
Moreover, such students advancing\textemdash without learning necessary concepts and skills\textemdash can become problematic in future settings. 
In this vein, developing means for assessing individuals’ contributions seems beneficial.
However, how to use them may vary situationally.
D2’s suggests that a measure of each team members’ contributions would be useful, but instructors should not make it the single factor in assigning grades to members. 

On the other hand, the design classroom is a community of practice, i.e., ``groups of people who share a concern or a passion for [design] and learn how to do it better as they interact regularly'' \cite{lave1991situated}.
In a community of practice, peripheral participation from some members is expected and legitimate. 
As these members gain understandings and skills, they take more central roles and contribute more. 
Learning science research has shown that in team settings, students who are less knowledgeable learn from those who know more \cite{webb1982student}. 

The present research contributes AI-based design analytics (Section \ref{sec:implications}) and computing them team member-wise has the potential to address the challenge of assessing contributions to team projects. 

\subsection{Frequent Feedback}
Students need specific and actionable feedback to understand where a design is lacking and make progress \cite{taylor2004juggling}.  
Instructors recognize that timely feedback is critical to student performance. 
At the same time, like prior investigations \cite{adams2016characterizing}, we found that instructors walk a fine line between interfering, which could hinder students’ ideation, and providing them with sufficiently frequent feedback to keep them on track and stimulate development.
Like Oh et al. \cite{oh2013theoretical}, we found that instructors provide feedback using a variety of modalities.
We also found that a common pain point among instructors is that they invest time into giving feedback, then often find that students fail to keep track of it and address it as they revise their designs.
More broadly, our study reveals students' lack of utilization of resources and incorporation of feedback provided to them.

\subsubsection{Feedback Modalities}
\begin{figure}[h]
  \centering
  \includegraphics[width=\linewidth]{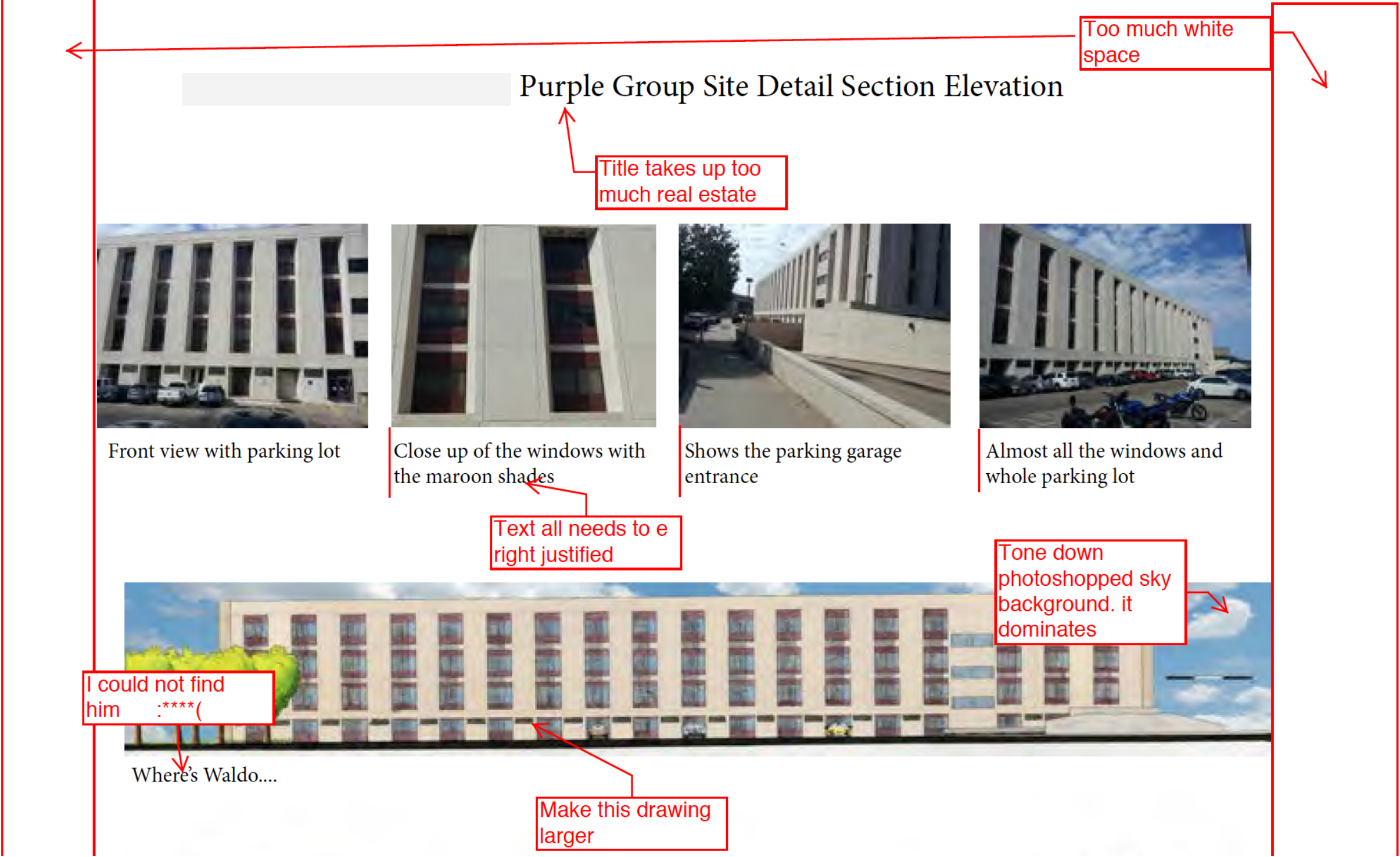}
  \caption{D6's feedback on a student's design through redlining. D6 marks up the problem areas and draws arrows connecting feedback written in respective boxes. The student receives feedback of using too much white space on the left and right sides, title taking more than needed space, text not styled per the instructor's guidelines, dominating background and small building size in the lower section, and an expected human figure.}
  \Description{Feedback through redlining.}
  ~\label{fig:redline}
\end{figure}

One common technique that design instructors use to provide feedback is to markup project documents. A salient form of markup is \emph{redlining} (Figure \ref{fig:redline}) \cite{jung1999immersive,brusasco2000computer,whyte2007visual}.
Practitioners either use a colored pen, typically red, to annotate printed materials, or do it electronically, with the help of computer tools. 
In conjunction with other forms of markup, such as sketches and annotations on tracing paper, the visual feedback helps designers iteratively formulate design problems and their solutions \cite{whyte2007visual}.

D9: \textit{We markup the documents ...and then give them a grade based on the rubrics that we have, along with the feedback. Feedback includes ‘something is missing’, ‘talk about something different here’...it may even be feedback that corrects some of their grammar.}

Given overall workloads, instructors sometimes face challenges in providing quantitative assessment. In order for students to understand their performance, instructors may provide verbal feedback. 

D4: \textit{I give them verbal feedback when they’re doing their presentation, but I don’t necessarily give them the complete grade scores right after their project, which they really want to know, I know that.}

The present research builds on these understandings\textemdash in conjunction with the understandings regarding students' lack of incorporation of feedback (presented next)\textemdash to derive new computational support that can assist instructors in their feedback processes.

\subsubsection{Students' Lack of Utilization of Resources and Incorporation of Feedback}
\label{sec:feedback_to_students}

In design courses, instructors often give students a variety of feedback opportunities. 
For example, D2 repeatedly says to her students that they can ask her to be a part of their ideation process and join their group meetings. 
Alas, she reports that students rarely contact her. 
D7 described that his course has 4 TAs, who are available to students 9-5 during the work week. 
This gives students the opportunity to consult with the TAs. 
He reports that they rarely do. 
Consequently, instructors are left with a sense that students could do better; if only.

Instructors sometimes ask TAs to regularly check student progress. 
This helps them become aware of issues, such as students who did not make enough progress, problems within a team, and lack of understanding of aspects of assignments. 
Balancing between interfering and guiding, instructors intervene when necessary. 
However, as D2 noted, \textit{“It creates another layer. It’s not just giving me all the information right away, so it takes some time.”} 
A problem for students and instructors is to maintain awareness of distributed information.

Another problem in student responses during design education is that the feedback does not get incorporated, even though iterative design is explicitly taught.
When discussing feedback processes, this problem was brought up by one of the participating instructors during the workshop.
Similar to the assessment of team contributions, this problem struck a chord with all participating instructors.

D11: \textit{I say [to students that] you have to fix something...In the programming studio [course], we'll say, you have to use less saturation. You cannot have all these things be saturated. And sometimes I'll end up saying it over and over before they fix it.}

D6: \textit{We have to kind of remind them, and then [our feedback] doesn't always all get incorporated. But that is how you get such polished products...that you continuously address some of [the feedback].}

Relatedly, we note that students often do not understand instructor feedback. 
Bridging this gap in understandings itself becomes an iterative process. 
Instructors point to the principles they taught, which students should demonstrate in their work as an evidence of meeting learning objectives.

D11: \textit{If there are some students that are struggling, then we can give them extra feedback...[Some] violate the principles that I teach them, and then I say hey, the principles, we talked about that. And they're like oh, you mean [that]...they get it.}

In the next section, we develop ideas for assisting instructors in providing feedback plus addressing needs for how to track feedback.

\section{Situating Analytics for AI}
\label{sec:implications}
How could analytics help solve these design education problems involving assessment and feedback?
Supporting design education with AI is challenging due to fundamental differences between the social and computational domains \cite{dourish2004we}. 
This interface border zone \cite{kerne2002interface}, where the fields intersect, is fundamentally sociotechnical.
Building on the fundamental work of Suchman \cite{suchman1987plans} and Dourish \cite{dourish2004we}, we find that traversing the social / AI border requires understanding situated practice.
The situated practices that we focus on take place in this border zone, involving instructors' design assessment and feedback. 
We build on these situated practices as a basis for discovering new approaches for deriving and presenting AI-based design creativity analytics.

In the user experience, analytics and how users interact with them, in the context of the task at hand, become inseparable.
Thus, we define \textit{situating analytics} as a methodology for conveying the meaning of measures that align with design rubrics, by contextually integrating the presentation of measures with associated design work.
Here, space of tasks at hand involves assessing and giving feedback on design in educational processes of project-based learning. 
Situating analytics is based on identifying and building with contextual properties, derived through understanding practices.
Interfaces traverse the border zone between the analytics and humans\textemdash here, students and instructors\textemdash making the analytics material \cite{giaccardi2015MaterialsExperience} in the human experience, i.e., ascribing meanings, mediating emotions, and supporting interpretation.

We use the situating analytics paradigm to develop ideas for integration of AI as \textit{algorithm-in-the-loop} \cite{green2019principles,green2019disparate} support for design assessment and feedback, toward validating outcomes of instructor and student interactions with AI as accurate, reliable, and fair.
As part of this, we develop ideas for a comprehensible and controllable AI integration, which can ``thereby [increase] the users' self-efficacy, leading to reliable... \& trustworthy systems'' \cite{shneiderman2020human}.

We invoke creative cognition's family resemblance principle as a basis for understanding the roles and limitations of AI-based analytics.
We then propose: (1) approaches for deriving AI-based design creativity analytics that align with and support instructors’ criteria; and (2) situating instructor and student interaction with the analytics\textemdash to foster transparency as well as validate assessment and feedback\textemdash through presentation, via dashboards integrated with design environments.

\subsection{Situating Analytics as a Paradigm for Conveying the Meaning of Measures}
\label{sec:situating_analytics_paradigm}
We offer our approach to situating analytics, for assessing creativity in project-based design learning, as a paradigm for making the meaning of measures transparent to users. 
To derive analytics, one AI approach is to extract features such as colors, number of words, and number of images, and then use a machine learning model to predict scores with high accuracy \cite{reinecke2013predicting}.
However, such an approach fails to provide actionable insight to users.
Instructors in our study highlight the need for approaches that can help students understand shortcomings in their work and give them insights about how to improve. 
Such assessments need to be learning objectives-based, in order to bridge the gap between instructors' and students' understandings.
Further, instructors expressed that they should be able to indicate whether AI is performing as expected.
These needs directly correspond to Shneiderman's principles: AI integration should be comprehensible and controllable \cite{shneiderman2020human}.

\begin{figure}[h]
  \centering
  \includegraphics[width=\linewidth]{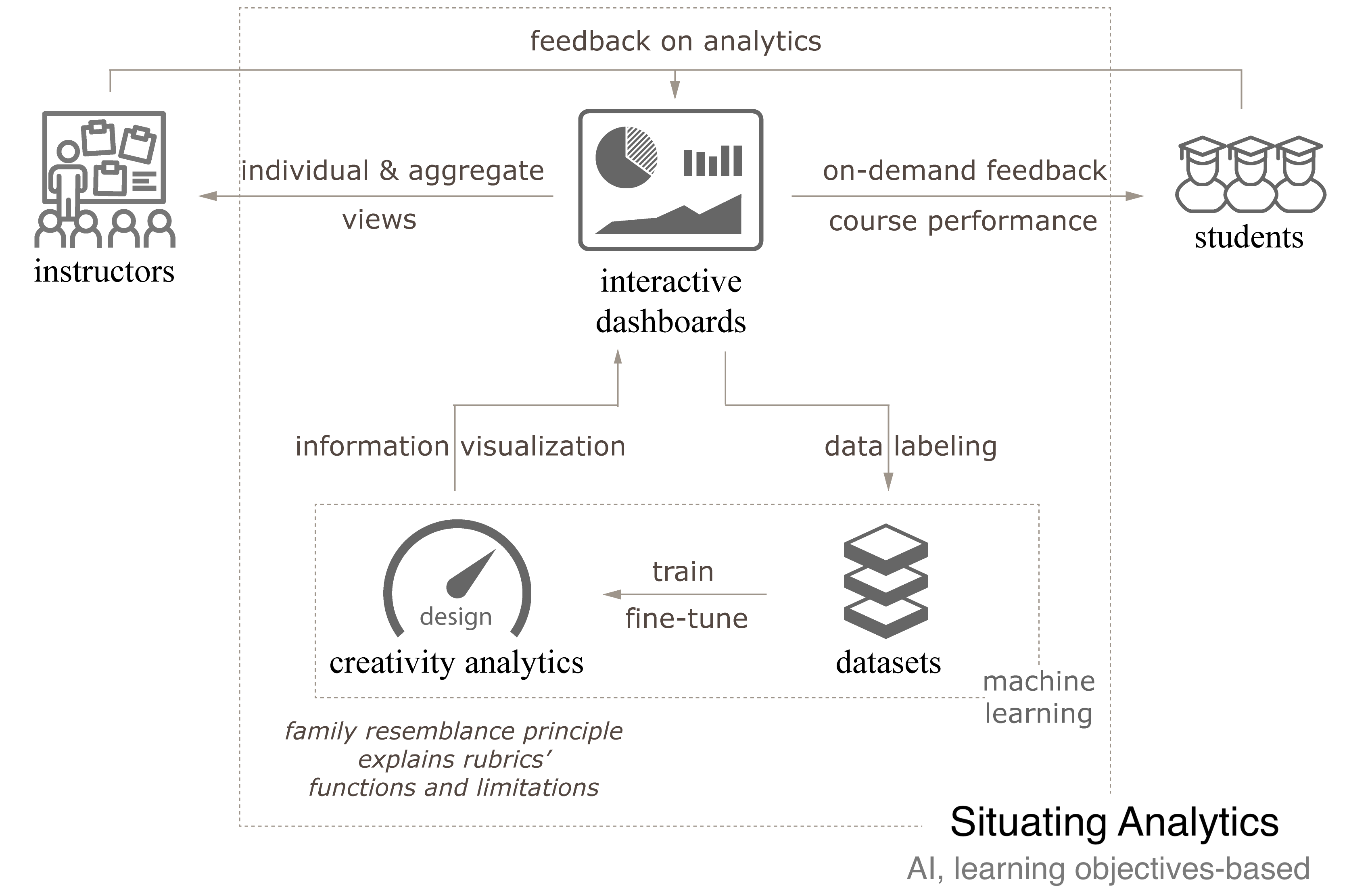}
  \caption{Diagram of a situating analytics approach to transparent design assessment and feedback. Transparent, learning objectives-based design creativity analytics\textemdash based on building with contextual properties derived through understanding practices\textemdash are presented via interactive dashboards to support teaching and learning, and simultaneously, to gather feedback for validating and refining analytics. (Use of icons under Creative Commons license. See attribution \cite{noun_project_icons}.)}
  \Description{Diagram of a situating analytics approach to transparent design assessment and feedback.}
  ~\label{fig:explaining_analytics_roadmap}
\end{figure}

D5: \textit{There's a kind of disconnect between [students] turning in a [design] and they getting a number [back]...Why is it a `B'?...[We] need to have a better tool communicating [the assessment] to the students.}

D6: \textit{Yeah that would be cool I think if we could develop some metrics to build into a consistent rubric. It will spit out the rubric scores and then the professor can say, well that's right or wrong.}

We diagram a situating analytics approach to transparent design assessment and feedback (see Figure \ref{fig:explaining_analytics_roadmap}). 
A co-design approach continuously involves instructor and student stakeholders in developing transparent, learning objectives-based analytics. 
Project-based learning analytics are derived using AI models that can constitute a basis for explaining the meaning of measures to the users.
Analytics are presented via dashboards integrated with learning environments.
Dashboard interaction can simultaneously support instructors and students in processes of teaching and learning, while at the same time gathering their feedback on the validity and utility of analytics. 
Dashboards need to present the approaches used to derive analytics, for intelligibility and accountability \cite{bellotti2001intelligibility}, in order to make how assessment works transparent. 
Through dashboard affordances, instructors and students will indicate whether and how an analytic and its derivation align with situated practices. They will articulate expected values and rationales when a computed analytic deviates. This data can be used to label analytics datasets that can be used to train supervised \cite{wang2015deepfont} and fine-tune unsupervised models \cite{jiang2017interactive}, as well as to in/validate analytics.

In Section \ref{sec:design_creativity_analytics}, situating analytics in practice, we prescribe AI approaches that can constitute a basis for conveying the meaning of design creativity analytics. 
In Section \ref{sec:situating_interaction}, we advocate dashboards integrated with design learning environments for situating instructor and student interaction with the analytics.

\subsection{Design Creativity Analytics}
\label{sec:design_creativity_analytics}
Design courses involve project-based learning. 
Creativity is vital. 
As Blikstein puts it, in project-based environments, there is a need for analytics that can assess “much more complex [characteristics], such as creativity [and] the ability to find solutions to ill-structured problems” \cite{blikstein2011using}. 
To measure creativity, its definition becomes essential. However, as discussed, established research avoids the epistemological trap of defining creativity in an absolute way (Section \ref{sec:sensitizing_creative_cognition}). 
Measuring creativity thus seems incompatible with most uses of learning analytics, which are to assess “specific and limited tasks” \cite{blikstein2011using}. 

This is where the \textit{family resemblance principle} \cite{smith1995creative} enables operationalizing design creativity analytics. 
Although creativity lacks an exact definition, according to family resemblance, specific characteristics of a design are likely to indicate creativity, even while no particular characteristic is necessary or sufficient (Section \ref{sec:sensitizing_creative_cognition}). 
In this way, we find that the family resemblance principle explains the rubrics that instructors specify for assignments.
Solutions are expected to exhibit characteristics, but not compulsorily. 
Extending the scope of this principle, design analytics have the potential to play the same incomplete, yet useful, role as design rubrics.

Thus, we prescribe situating design creativity analytics that indicate the likelihood of quality design, without being absolutely definitive. 
We argue against any one computed analytic being seen as necessary or sufficient. 
Rather, we advocate \textit{suites of design analytics}, each with the potential to measure characteristics that often occur in good design solutions. 
\textit{Design analytics suites need not be complete, in order to add value.}

We note that situating analytics corresponding to instructors’ every rubric element may not be possible. 
For example, it can be challenging to computationally assess \textit{“a spatial sequence [in an architecture] that leads [a person] to an intense highlight moment (D1)”}. 
Our goal here is to augment, not replace instructors’ roles.

The instructors in our study believe that some feedback about design, which they now provide “by hand”, can be computed\textemdash and provided on demand\textemdash in order to help students to more quickly understand problems and correct them. 
We argue that AI techniques are suitable for various design characteristics, providing means to compute on demand, transparent, learning objectives-based analytics.
Further, consistent with ongoing forays into process-oriented creative work \cite{lippard1997six,kaprow2014happenings,higgins2002fluxus,seitz1961art}, we found that design process is important in how instructors assess student work. 
Thus, a derivation of analytics would beneficially incorporate process characteristics. 

We develop a set of situated \emph{design creativity analytics}, to computationally derive actionable measures for giving students feedback on project-based learning work.
Each analytic is mapped, by family resemblance, to a  criterion, which we discovered in design education rubrics, across situated contexts of our study (Table \ref{tab:participants}).
Some design creativity analytics correspond to previous creative cognition ideation metrics \cite{shah2003metrics,kerne2014using,jain2015tweetbubble}: \emph{Fluency}, or the number of ideas (D2, D8, D9); and \emph{Flexibility}, or the diversity of ideas (D8, D9, D11).
Other design creativity analytics are, as far as we are aware, new to creative cognition: \emph{Visual Consistency} (D1, D6, D11), or the presentation of similar information using similar attributes; \emph{Multiscale Organization} (D6, D11), or the presentation of ideas as a hierarchy; and \emph{Legible Contrast} (D5, D6, D11), or the juxtaposition of hues for legible presentation.
To further distinguish, a prior creative cognition ideation metric, Visual Presentation \cite{kerne2014using}, assesses aspects of Visual Consistency, but it does not assess Multiscale Organization and Legible Contrast.
Appendix \ref{sec:design_characteristics} enumerates additional potential learning analytics, which we identified, and find relevant to design education.

D6:\textit{...[the computer] could look and say you only got 2 drawings at the [larger] scale where you got 35 at the [smaller] scale, [hence] you need to do more analysis of the larger scale.}

D11: \textit{...[the computer could tell] where high contrast is vs. where there is less contrast. [This matters] because contrast takes human visual attention.}

\subsubsection{Fluency}
Instructors in our study use the number of ideas\textemdash known as the Fluency ideation metric in prior creativity research \cite{shah2003metrics,kerne2014using}\textemdash as a key analytic for assessing creative design.
For example, D8 requires students to brainstorm and come up with as many ideas as possible in the beginning phases of design, stating that the activity's ``goal is quantity''.
According to Darwinian theories of creativity, more the number of ideas, higher is the likelihood that one of them will survive and grow to be creative \cite{guilford1950,simonton1999}.
In the brainstorming process, students develop ideas in textual as well as visual forms.
Language and imagery represent complementary processes of human cognition \cite{baddeley1992working} and their combination is known to aid the formation of mental models \cite{glenberg1992comprehension,mayer2003nine}. 

Prior creativity research developed computational means of assessing Fluency, such as the number of text \cite{kerne2014using}, image \cite{kerne2014using}, and sketch \cite{kerne2017strategies} elements that comprise a free-form visual semantic composition of ideas.
A recent investigation found that Fluency assessments provide instructors with first-order insights into student effort levels on design projects \cite{britain2020}. 
The paper reported that instructors desire an advanced analysis of design work, such as extraction of ideas contained within text and image elements.

To compute design creativity Fluency analytics, researchers can use state-of-the-art AI-based content recognition\textemdash e.g., natural language processing (NLP) \cite{young2018recent} and computer vision models \cite{ordonez2011im2text}\textemdash to extract ideas contained within text, image, and sketch elements.
An example is Mackeprang et al.'s investigation for supporting collaborative design, where they use NLP models to extract ideas contained within participants' text entries \cite{mackeprang2018concept}.
As computer vision models have been found to generate highly accurate image \cite{ordonez2011im2text} and sketch \cite{yu2017sketch} descriptions, researchers can utilize them to extend Mackeprang's approach to designs that compose diverse types of elements, such as text, image, sketch, video, and embedded docs and maps.

\subsubsection{Flexibility}
We invoke the ideation metric Flexibility/Variety from creative cognition \cite{shah2003metrics,kerne2014using} to represent instructors' use of the number of categories of ideas as an analytic for assessing design work.
For example, D9 requires students to “organize [generated] ideas into concepts”, and then include “at least four ‘different’ concepts” and “a discussion of the advantages and disadvantages of each concept”. 
In design creativity research, Flexibility/Variety represents the span of the explored solution space \cite{smith1995creative,kerne2014using}.
Flexibility measures opportunities for remote associations, which are vital in developing creative solutions \cite{mednick1962associative}.

Prior creativity research developed computational means for assessing Flexibility, such as the number of different webpages, websites, and website types that the users collected ideas from \cite{kerne2014using}.
Recent work on collaborative design ideation examined the use of external knowledge graphs such as Wikidata and DBpedia to compute semantic distances among ideas, toward assessing Flexibility/Variety \cite{mackeprang2018concept}.
Researchers have used computational methods to organize ideas into a tree structure. 
Examples include Linsey et al.'s use of WordNet \cite{linsey2009increasing}, Fu et al.'s latent semantic analysis in conjunction with a posterior probability based method \cite{fu2013meaning}, and Vattam et al.'s use of functional hierarchies \cite{vattam2011dane}.
The tree structure supports analogical thinking and has been used to assess semantic distances among ideas.
A promising approach is to investigate vector word representations to compute semantic distances \cite{pennington2014glove}.

To compute design creativity Flexibility analytics, first, contextualized vector representations of words \cite{pennington2014glove} can be created using text sources, such as books, scholarly articles, and patents in a design field.
As noted above, AI-based computational models can be used to extract ideas from text, image, and sketch elements of a design.
Then, using the vector word representations, semantic distances among these ideas can be calculated \cite{pennington2014glove}.  
Categorizing ideas, based on these semantic distances, has the potential to provide measures that correspond to instructors’ understandings and assessments.
Graph visualization of ideas, based on semantic distance, has the potential to support reflection. Walking across these graphs has been shown to support analogy formation \cite{gentner1983structure, linsey2009increasing}.

\subsubsection{Visual Consistency}
Visual Consistency refers to using graphical attributes\textemdash such as size, color, and font\textemdash in similar ways for presenting similar types of information. 
For example, all section headings could be in boldface, 14-pt and all subsections headings in italic, 12-pt Helvetica font. 
Instructors teach students to follow principles of visual design, through which even complex information can be presented with clarity \cite{tufte1990envisioning,bertin1983semiology}.
D11 provides guiding instructions to students for developing a \textit{visual program}\textemdash which includes “a grid structure, consistent type sizes and styles, and a color plan”\textemdash based on Meggs’ definition: “a system of parameters used consistently to unify a series or sequence of designs” \cite{meggs1992type}.

Prior creativity research lacks computational methods for assessing consistency.
Human raters assigned scores for the use of visual design principles such as whitespace, arrangement, and coherence, as part of assessing Visual Presentation creative cognition ideation metric \cite{kerne2014using}.

To compute design creativity Visual Consistency analytics, in cases where a design schema stores information about the type of each element (e.g., heading, subheading, caption), the attribute values for elements of the same type can be compared. 
Inconsistencies can be highlighted. 
Alas, such schematized feedback is absent from typical design environments, e.g., Photoshop, Illustrator, and InDesign. 
In such cases, elements could first be clustered using attributes such as position and size \cite{jain2021recognizing}, as well as ideas \cite{jain2016towards} contained in them. 
Then, if two clusters have similar relative positioning of elements within them, then the attributes of the corresponding elements can be compared, to highlight inconsistencies.
Further, the number of clusters can indicate the use of whitespace for organizing information.

Recently, clustering algorithms based on users’ implicit actions of organizing content in a 2D design space have shown promising results \cite{siangliulue2016ideahound}.
Design environments can utilize such algorithms, when process data is saved.
For example, they can incorporate user actions of selecting, moving, and manipulating elements together within the design space toward determining spatial proximity.

\subsubsection{Multiscale Organization}
Multiscale Organization refers to the visual and conceptual representation of ideas using hierarchy. 
It is a foundational element in design \cite{barba2019cognitive}.
D6 requires students to visualize their landscape architecture project data ``from national to regional to site scale''. 
D11 engages students in multiscale organization, through a collaborative, zoomable design space and guides students to ``use scale to nest sets of elements, where appropriate, to create readability, since [they] are sure to have more elements than can fit on screen, and in human working memory, at one time''.

Multiscale organization supports designers in exploring, juxtaposing, and synthesizing ideas and their relationships across multiple scales \cite{lupfer2018multiscale}.
It allows them to connect with and develop a design, by shifting their cognitive point of view to different scales or levels \cite{barba2019cognitive}.
Design environments supporting multiscale organization\textemdash e.g., Photoshop, Illustrator, and IdeaMâché \cite{lupfer2019multiscale,lupfer2016patterns}\textemdash allow going beyond 2D and assembling content at multiple zoom levels or scales.
Lupfer et al. found that, for students using IdeaMâché to perform free-form web curation, multiscale organization supported students' iterative and reflective ideation processes when working on design projects \cite{lupfer2019multiscale}.
 
To compute Multiscale Organization analytics, as a starting point, the number of scales or zoom levels at which students have organized their ideas can be computed \cite{jain2021recognizing,jain2017measuring}.
Then, the assessment can focus on consistency aspects discussed in the last section, i.e., compare clusters scale-wise to find whether similar information is presented using similar attributes.
For multiscale clustering of student design, researchers can investigate AMOEBA and its extensions, which use Delaunay triangulation to compare distances among different elements and recursively determine spatially nested groups \cite{estivill2002multi,jain2021recognizing}.
They can likewise investigate identifying nested groups using Self-Organizing Map (SOM) techniques, which preserve topology and have proven effective for spatial clustering \cite{baccao2005self}.

Further, techniques based on process data \cite{siangliulue2016ideahound} can be extended to multiscale clustering algorithms.
For example, user actions of zooming in or out, followed by the selection and manipulation of a set of elements, can be used as a factor in determining different scales of content organization.

\subsubsection{Legible Contrast}
Contrast refers to color properties that can be juxtaposed to produce a range of visual effects \cite{itten1970elements}.
D11 specifies in assignment: “...appropriate use (not too much!) of contrast”. 
This characteristic is based on visual design principles articulated by Tufte \cite{tufte1990envisioning}, which have been used by prior creativity research for assessing visual presentation \cite{kerne2014using}.
In regard to contrast, Tufte invokes Imhof’s first rule of color composition: large adjacent areas of pure, bright colors are loud and unbearable, but when used sparingly, can help achieve extraordinary effects \cite{windisch1941schule,tufte1990envisioning}. 
In Figure \ref{fig:redline}, such adjacent areas reduces the focus on the building in the foreground.
As D6 redlined, ``Tone down photoshopped sky background. It dominates''.

To compute design creativity Legible Contrast analytics, image processing and computer vision algorithms that are capable of identifying regions of high contrast \cite{osberger1998automatic} can be utilized.  
Such algorithms operate on extremely small blocks, so one potential way is to find the percentage of such blocks in an overall design. 
Another possibility is to detect thick lines or boxes by finding long sequences of these high contrast blocks. 
For smaller occurrences, an interactive machine learning approach \cite{jiang2017interactive} would be beneficial for obtaining instructor feedback, to iteratively improve identification of whether the use of high contrast is excessive in the context of a design. 

As a large collection of examples with such feedback becomes ready, deep learning and domain adaptation techniques \cite{wang2015deepfont} can be explored. 
As we discuss below, a learning analytics dashboard can play a new role in this collection of labeled examples.

\subsection{Situating Interaction with Design Analytics}
\label{sec:situating_interaction}
Enhancing sociocultural contexts with AI is challenging due to the multitude of cognitive, affective, and latent factors in play \cite{rikakis2018progressive}.
The statistical intelligence of AI/ML algorithms may provide an interpretation that drastically differs from common sense human interpretation of data \cite{dove2017ux}.
While algorithm-in-the-loop AI approaches have been shown to improve accuracy in multiple contexts, they are prone to bias and errors \cite{green2019principles}.
A successful human-AI system requires ``careful design of the fine structure of interaction'' that makes AI integration \textit{controllable} and \textit{comprehensible}, which thereby increases users' reliability and trust in the system \cite{shneiderman2020human}.

We propose integrating analytics dashboards with design learning environments, as a means for jointly (1) situating human-AI interactions within the contexts of design courses; and (2) directly meeting instructors’ and students’ practices and needs.
As part of situating instructor and student interactions with dashboards, we develop ideas for affordances through which users can in/validate analytics and provide feedback that can be used to improve AI algorithms.
Through analytics dashboards integrated with design, instructor processes of providing feedback and students needs for receiving such feedback have the potential to be functionally rendered as isomorphic with AI algorithms' iterative needs for feedback on examples, in order to constitute labeled datasets for training recognizers.
That is, we call for analytics dashboards integrated with design, in order to pair instructors giving feedback and students receiving it, with feedback on AI-computed design creativity analytics.

Turnbull explains indexicality as involving maps over space and time, which convey information that ``can only be [completely] understood within the ... specifics of the circumstances and cannot be generalized beyond that context'' \cite{turnbull1993maps}.
We use this idea to motivate the need for connecting design analytics dashboards with learning environments.
Design analytics dashboards are indexical, i.e., they derive and present analytics based on characteristics that exist and can best be understood within the context of actual design work \cite{jain2024indexing}.
Indexically connecting AI-based design creativity analytics with design elements in learning environments improves comprehensibility.
Such indexical connections support users in understanding the basis of analytics and in/validating them, and through this process, providing feedback that AI designers can utilize to improve recognition.

\subsubsection{Design Analytics Dashboards}
In our study, instructors brainstormed ways in which analytics \cite{siemens2011penetrating} and visualizations \cite{card1999readings} could help them gain insight into students’ performance. 
They expressed needs for tracking design processes across multiple levels. 
For example, D2 would like to be able to compare points a student got last week, or on previous projects, to find out if the student has improved. 
D4 imagined visualizations that would reduce time and effort to make reports at the end of a semester, or at the end of the project, so that there isn’t a need to \textit{``write every single thing from scratch''}. 
D6 wants to see measures of how an individual student or team worked with an artifact (process scores / analytics), coupled with characteristics of that artifact product.

D6: \textit{Is there a way...we can keep track of whether or not they got better presentation scores, based on how many times they practice?...We can see where they fixed [their design] before they actually come in and present.}

Dashboard, defined as ``a visual display of the most important information needed to achieve one or more objectives'' \cite{few2013information} can be useful in these regards.  
The tabular format\textemdash each row presenting an observation, with columns corresponding to attributes\textemdash supports ``flexible querying and many perspectives for data exploration'' \cite{yalccin2017keshif}.
Prior learning analytics dashboards \cite{duval2011attention} have facilitated quick understanding of the progress of student teams and individuals. 

However, their efficacy has mostly been investigated in lecture-based learning contexts. 
Using dashboards for project-based learning requires personalization \cite{michel2017supporting}.
As design assessment lacks absolute criteria, design analytics dashboards specifically need to enable each instructor to select and combine characteristics, based on their pedagogic orientation and the project at hand.

In design education contexts, AI-based design creativity analytics will measure complex characteristics.
As Bellamy et al. discuss, \emph{bias} can result from systematic error in AI training data or models; bias can produce unfair outcomes for individuals or groups within a population \cite{bellamy2018ai}.
Greene et al.'s algorithm-in-the-loop research advocates for users' rights to challenge assessments that affect them, as AI algorithms often fail to adapt to novel circumstances \cite{green2019principles}. 
According to Woodruff, users' ability to challenge and change AI decisions\textemdash thus allowing contestability and recourse\textemdash are approaches for addressing algorithmic bias \cite{woodruff201910things}.
Further, to design AI technology that users can trust, Woodruff et al. discuss the need for strong participation of stakeholders and consideration of multiple perspectives \cite{woodruff2018qualitative}.

Design analytics dashboards have the potential to situate instructor and student interaction in forms that mitigate complex problems posed by the use of AI.
Dashboards can actively engage instructor and student stakeholders in continuous validation and refinement of AI models and derived design analytics. 
Dashboards can concurrently make analytics available in design learning contexts and gather data about how the analytics, as well as their presentation, affect design learning experiences \cite{jain2024indexing}.
In situating stakeholders' interaction, comprehensibility and controllability will be vital.

\textit{Comprehensibility}: In Section \ref{sec:feedback_to_students}, we described how explaining the basis of assessment to students is important in helping them improve their work.
As D11 expressed, \textit{``[Some] violate the principles that I teach them, and then I say hey, the principles, we talked about that. And they're like oh, you mean [that]’’}.
Analogously, instructors expressed the need for AI-based analytics to be comprehensible. 
Recalling D5, (Section \ref{sec:situating_analytics_paradigm}): \textit{``There’s a kind of disconnect between [students] turning in a [design] and they getting a number [back from AI]...why is it a `B'?''}.

\begin{figure}[H]
  \centering
  \includegraphics[width=0.98\linewidth]{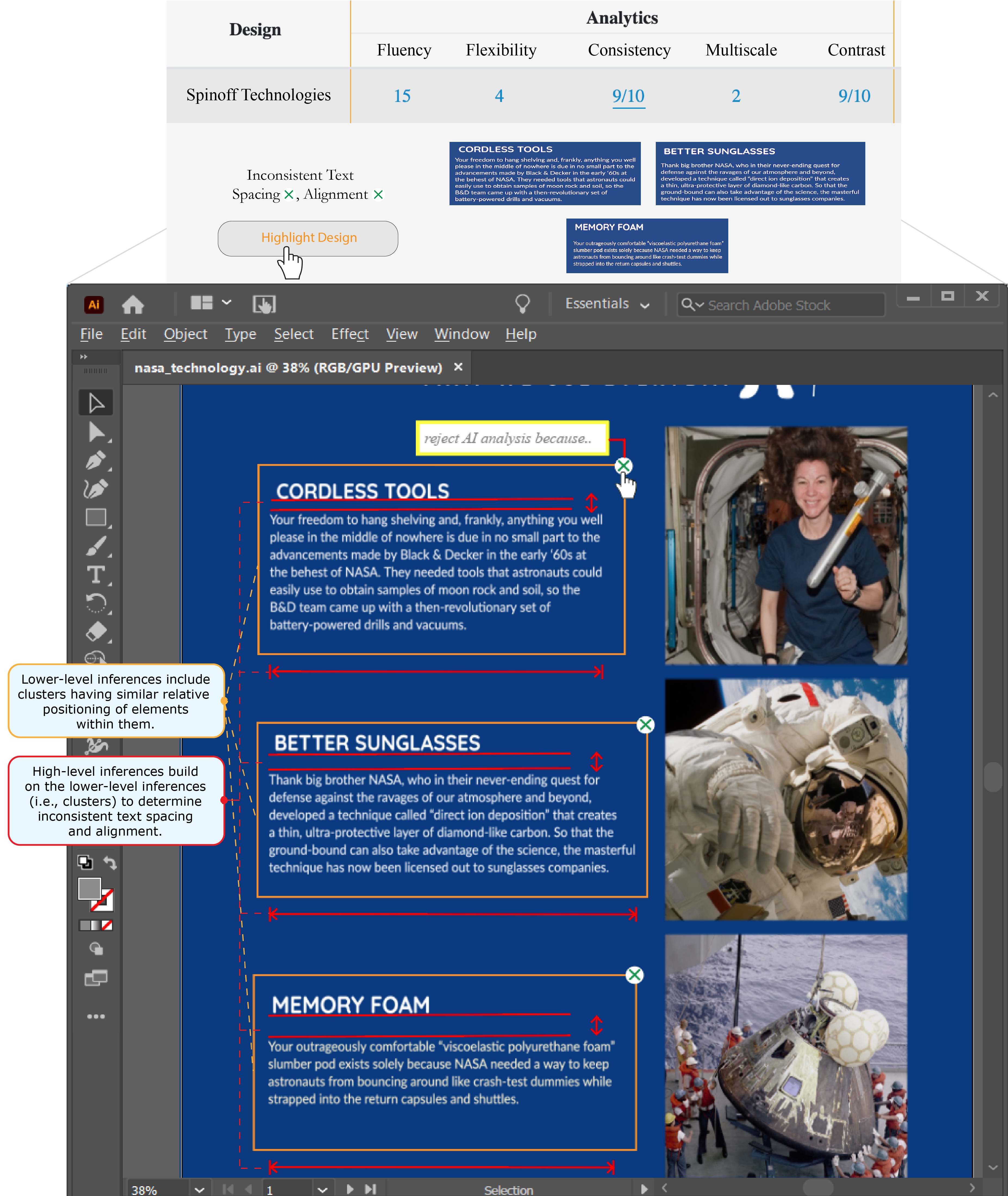}
  \caption{A mockup of dashboard integration with design learning environments. Design analytics dashboards are indexical: they present analytics that refer to and can be understood more effectively within the context of actual design work. For a quick overview, activating an analytic can present its underlying basis, right there on the dashboard. To aid \textit{comprehensibility}, on activating the presented information, dashboards can support highlighting/redlining the elements\textemdash that form the basis for analytic computation\textemdash within the actual design work. To aid \textit{controllability}, the information presented within dashboards and learning environments can include affordances for users to challenge the assessment and rationale behind it. This mock-up shows clusters (in orange) used as the basis for Visual Consistency analytics and inconsistencies identified (in red). Users can challenge assessment (activating cross symbols) and input rationale (in the presented text box). See attribution for `NASA Technology' design example \cite{adobeSparkTemplate}.}
  ~\label{fig:mockup}
\end{figure}

To aid comprehensibility, dashboards need to explain (make visible) the underlying basis for computing analytics. 
For \textit{Fluency} analytics, dashboards can present ideas extracted from text, image, multimedia, and sketch elements.
For \textit{Flexibility} analytics, dashboards can relate extracted ideas by visualizing semantic distances among them.
For \textit{Visual Consistency} analytics, dashboards can identify elements that use dissimilar attributes despite being of the same type (Figure \ref{fig:mockup}).
For \textit{Multiscale Organization} analytics, similarly, dashboards can identify inconsistent elements scale-wise.
For \textit{Legible Contrast} analytics, dashboards can present regions of high contrast.
Connecting the presentation with learning environment can further aid in comprehensibility (Section \ref{sec:connecting_environment}).

We note that our approach of explaining analytics involves making the lower-level inferences that AI makes transparent.
These include various conceptual and visual aspects present in the design (as detailed in Section \ref{sec:design_creativity_analytics}).
Examples of conceptual lower-level inferences include ideas extracted from text, image, and sketch elements.
Examples of visual lower-level inferences include clusters \cite{jain2021recognizing} and areas of high contrast present in the design.
High-level inferences are the analytics\textemdash Fluency, Flexibility, Visual Consistency, Multiscale Organization, and Legible Contrast\textemdash derived using these lower-level inferences.
To maintain focus, the present research only develops approaches for explaining how analytics are derived using lower-level inferences, not the lower-level inferences themselves, and so on.
For example, for comprehensibility and controllability of Fluency analytics, we present to the user the ideas considered toward computing the Fluency, but not how AI determined whether a given image is of a cat or a dog.

\textit{Controllability}: In case the AI assessment does not make sense, instructors want outcomes to be controllable. 
As D6 expressed (Section \ref{sec:situating_analytics_paradigm}), \textit{``the professor can say, well [the assessment is] right or wrong''}.
The presentation of the underlying basis allows instructors and students to challenge not only an analytics score, but also the approach used to compute it.
To aid controllability, dashboards can include affordances for instructors and students to express if and how a given design analytic makes sense. 

Human-AI interaction through dashboards can also be structured as a form of labeling data, to enable instructors to simultaneously give feedback to students and developers, and about datasets. 
For example, if instructors disagree with any computed score and/or approach, they can be presented with affordances to enter an alternative score and the rationale behind it. 
With an increasing number of examples, machine learning algorithms\textemdash including deep learning with attention mechanisms \cite{hudson2018compositional}\textemdash can be investigated to more closely model instructors’ rationale and assign scores for a new design in the given context. 
Further, using transfer learning \cite{pan2009survey}, the data can potentially be utilized in other course contexts.

In the context of situating instructor and student interaction via dashboards, we note that the purpose is to make AI integration controllable and comprehensible, as an approach for addressing algorithmic efficacy and bias. 
We are not advocating poorly performing AI algorithms, which can increase the load on students and instructors, by requiring extensive input from them when algorithms produce incorrect results.

\subsubsection{Integrating Dashboards with Design Environments}
\label{sec:connecting_environment}
As described above, dashboards are indexical: they present design creativity analytics\textemdash based on characteristics valuable to the instructor and student users\textemdash which refer to and can be understood, more completely, within the context of the actual design work \cite{jain2024indexing}.
The indexicality of the dashboard functions as a mediating mechanism between AI-based analytics and design work.
This integration improves the comprehensibility of analytics by connecting them to corresponding design elements in learning environments.

For \textit{Fluency} and \textit{Flexibility} design creativity analytics, if the user wants to understand the spatial and conceptual contexts in which an extracted idea occurs, a dashboard can support activating the idea and highlighting its occurrences within a design.
The user could then indicate whether the AI-algorithm correctly extracted an idea from respective spatial and conceptual contexts, and input if a different idea would be more suitable.
For \textit{Visual Consistency} and \textit{Multiscale Organization}, likewise, if the user wants to see the basis of comparison of attributes of two elements, a dashboard can support redlining inconsistencies and presenting\textemdash within the design\textemdash the scales and clusters based on which different sets of elements are being compared to each other (Figure \ref{fig:mockup}).
The user can then indicate whether the AI-algorithm correctly identified scales and clusters. 
For \textit{Legible Contrast}, a dashboard can likewise support redlining the regions determined as high contrast and viewing them within the design, in the context of surrounding elements, to allow the user in validating whether the use of contrast is indeed excessive.

\textit{Tracking Feedback}: Integrating dashboards with design can help students and instructors track feedback.
Further above, we discussed students' lack of incorporation of feedback, despite reminders, is a common pain point for the instructors in our study.
Instructors find it frustrating, as they put significant efforts into providing frequent feedback.
Verbal critique and physical redlining of physical copies are often lost and feedback provided via learning management systems \cite{beatty2006faculty} also does not get incorporated.
The problems that instructors point out continue to exhibit across iterations.

Firstly, to enable connecting verbal and physical feedback with design elements, in learning environments, computational support needs to facilitate digitization.
Secondly, the needs for connecting and tracking immediately suggest a combination of node-link \cite{berners1994world,nelson1990xanadu}, compound \cite{nelson1993literary}, and spatial hypertext \cite{marshall1995spatial} approaches.
Using these approaches as the basis, dashboards can visualize where the feedback originated and how the design changed over iterations, including feedback incorporation, or failure to incorporate, to varying extents.
Design versioning features such as visual ``diffs'' \cite{gitForDesigners} will facilitate making design feedback and iteration visible. 

Direct support for redlining \cite{jung1999immersive,jung2000immersive} students’ assignments, within digital environments of design tools\textemdash and creating connected dashboard representations\textemdash also needs to be investigated.
Redlining can be added, as a special CSCW layer, in tools such as Photoshop and InDesign. 
In conjunction with community feedback on creative design \cite{kim2017mosaic}, this design process integration has the potential to transform the processes of teaching and learning. 
As students iterate on design work, they can respond using this layer, which will trigger notifications to instructors that students have addressed particular feedback.
In addition, the layer can support instructors' needs to flag the presence of feedback and if it has been addressed.

D6: \textit{...if students could just show their work and we could just draw straight on it. That would help us a lot. It’s instant feedback, you know.} 

D2: \textit{Yeah, and also sometimes if I can kind of put a flag, and give me a reminder if this person [incorporated the particular feedback].}

By identifying commonalities, involving feedback across deliverables, recurrent problems from this layer can be highlighted to instructors and students on dashboards. 
This information can stimulate instructors’ pedagogical intervention and students’ seeking helpful resources. 
The problems can be categorized at student, team, and course levels for effective understanding.

\section{Co-Design: Build Stakeholder Community to Support with AI}
\label{sec:pd_building_community}

We advocate using co-design methods for investigating AI support for human and social computing contexts. 
These methods actively involve users across stages of design processes \cite{sanders2012convivial,sanders2008co,kyng1991design}.
The methods situate users as the ‘experts of their experience’ \cite{sanders2012convivial}. 
Their expertise involves forms of tacit and shared knowledge and communication \cite{schuler1993participatory}.
Making the tacit \textit{visible} plays a crucial role in designing support for participants' work practices \cite{suchman1995making}. 
Co-design provides designers with methods that support users in expressing and sharing their expertise \cite{sanders2012convivial}.

We actively involved our “users” through co-design approaches. 
Their stakes in the process and levels of engagement increased. They valued the work more. They become more forthcoming and more available.
Participating instructors shared ideas on addressing common pain points. 
Findings and implications are derived \textit{with} them, not \textit{for} them. 
Design with stakeholders, rather than for stakeholders, deepens involvement \cite{disalvo2014designing}.

As described in the methodology (Section \ref{sec:methodology}), two instructors chose to participate further, e.g., in writing this paper. 
Participating instructors expressed interest, and began adopting a design ideation environment \cite{hamilton2018collaborative} that the initial research team is developing.
They have begun using assessment and feedback support capabilities designed in conjunction with ideas presented in the paper.
The PI and one of the participating instructors, who became a co-author, together published a work-in-progress based on user experiences with an initial version of design creativity analytics and dashboards in course contexts.
Further, the PI and two participating instructors who became co-authors have started writing grant proposals together, addressing design education challenges, identified across course contexts, through the co-design process.

Learning analytics and dashboard environments that represent them to users, function as boundary objects \cite{star1989institutional}: “physical or conceptual entities that each [stakeholder] interprets in its own way, but that provide common referents or points of articulation to ground conversations” \cite{suthers2013learning}.  
These boundary objects, in conjunction with co-design methods, can facilitate addressing ``the `middle space' where learning and analytics meet'' \cite{suthers2013learning}, and answering questions such as whether analytic computations provide actionable insights to stakeholders.
In the course contexts of our study, design creativity analytics, presented via dashboards integrated with learning environments, have started providing common referents for instructors, students, and developers of AI-based computational support.
In our initial observations, a triumvirate of situated artifacts\textemdash design rubrics, AI-based analytics, and dashboards that enable controlling and comprehending outcomes\textemdash support grounding conversations among stakeholders. 
Further research is needed.

Stakes and engagement evolve through co-design, to manifest growing interest, usefulness, and involvement. We envision the community growing further, while providing new opportunities for multidisciplinary design discourse and collaboration.

\section{Conclusion}
We developed a case study for human-centered AI, engaging aspects of co-design with instructors across diverse fields and developing understandings of situated practices, in order to investigate whether and, if so, how new forms of computation, e.g., AI, could support teaching and assessing design.
Our findings are affirmative.
Using a grounded theory approach to analyze data from workshops and discussions, as well as design artifacts from courses, we formulated categories and presented relevant themes, focusing on: 1) assessment and feedback challenges and 2) implications for AI-based analytics.

Our case study contributes new theory: 
(1) understanding of uses and limitations of rubrics through creative cognition’s family resemblance principle;
(2) \emph{situating analytics}, as a paradigm for for conveying the meaning of measures that align with design rubrics, by contextually integrating the presentation of measures with associated design work;
(3) operationalizing design creativity analytics, based on family resemblances in design rubrics, which has the potential to provide students with actionable, on demand feedback, and thus support them in learning to do design through iterating on their design work; and
(4) dashboards, integrated with design environments, for situating instructor and student algorithm-in-the-loop interaction with AI-based analytics.   

We invoke the family resemblance principle to contribute new understanding of how rubrics work, and likewise for how AI-based design creativity analytics, derived from process and product data, can productively be incorporated into education. 
Family resemblance tells us that no particular characteristic is essential, but together\textemdash in a rubric or analytics dashboard\textemdash they tend to indicate good design work. 
As D5 expressed, one student could do something really simple, while another something really complex; both may produce good designs. 
Hence, while rubrics and analytics provide vantage points, they do not provide a God’s eye view \cite{sankey2003scientific}.
Analytics’ purpose is to augment, not replace, instructors’ ongoing interpretation and engagement.

We lay out \textit{situating analytics} as a paradigm for conveying the meaning of measures to the users.
In the present research, this means for assessing design in educational processes of project-based learning. 
Goals are to bridge comprehensibility, controllability, and actionability gaps between users and AI.
Gaps between users and AI have been recognized in the healthcare domain, where despite success in lab settings, AI-based clinical decision support tools sometimes fail in practice, due to, ``a lack of consideration for clinicians' workflow'' \cite{yang2019unremarkable}.
Through a situated, co-design approach, we discovered characteristics that instructors seek when assessing design. 
Next, based on these contextual properties, we identified analytics corresponding to salient conceptual and visual characteristics, such as Fluency, Flexibility, Visual Consistency, Legible Contrast, and Multiscale Organization of ideas.
Then, for deriving these analytics, we focused on AI-based approaches that can constitute a basis for conveying the meaning of measures to the users.

We argue that explaining AI outcomes is vital to human-centered AI;
it will make AI more valuable to people and society.
A `father' of modern AI tweeted, \textit{``Suppose you have cancer and you have to choose between a black box AI surgeon that cannot explain how it works but has a 90\% cure rate and a human surgeon with an 80\% cure rate. Do you want the AI surgeon to be illegal?''} \cite{hinton2020tweet}.
In response, there was a huge debate. Many opposed the black box AI, arguing that the 90\% rate system might not have been trained on a representative sample and the system cannot explain or make that transparent to the patient.
In the present research, without explaining analytics, students would fail to understand why they are assigned certain scores, and especially, how they can improve their designs.
Instructors would face challenges in relating analytics to design processes.

One common pain point for instructors in our study is assessing contribution in team projects.
Further research can investigate computing design creativity analytics, team member-wise, to understand efficacies of providing instructors and students with insights that are complementary to teamwork assessment\textemdash e.g., peer ratings on contribution to work, interaction with others, and skills\textemdash using tools such as CATME \cite{loughry2014assessing}.

Kerne identified the interface as an integrated conceptual and sensory border zone, which supports interplay among humans and technologies \cite{kerne2001collagemachine}.
To address challenges of traversing social / AI border zones for supporting design course contexts, we developed ideas for how to give design analytics dashboards participatory, multi-function roles in algorithm-in-the-loop assessment.
Dashboards that situate presentation of and interaction with analytics have the potential to stimulate instructors’ intervention and students’ continuous improvement of work.
The dashboards are expected to help students, by giving them new forms of on demand, transparent feedback on their design work. 
The dashboards will help instructors, by giving them new views, both individual and aggregate, which provide insights about what is going on in their classes, in terms of how students are accomplishing and failing.
In turn, instructors can use these dashboard views to formulate new plans for what students need.

In addition to meeting instructor and student practices and needs, dashboards can simultaneously situate their algorithm-in-the-loop interactions to iteratively refine and validate derived analytics. 
By providing these stakeholders with affordances for indicating whether or not an analytic and its derivation make sense, and input of expected value and rationale when AI does not match, dashboards have the potential to help generate and refine labeled analytics datasets, and so make models work better.

To increase the intelligibility of analytics, we propose integrating dashboard presentations with design learning environments.
Connecting the instructor, as well as AI assessment and feedback, with design elements of concern can help address problems of how students lose track of feedback, another common pain point for instructors in our study. 
We further developed ideas involving digital redlining over student work, as a special feedback layer, in various design tools, to assist with tracking.

Situating instructor and student interaction with analytics\textemdash by integrating dashboards with actual design work, in design environments\textemdash has the potential to transform design education. 
To keep pace, tools such as Photoshop, Illustrator, and Sketch need to incorporate support for learning.
We expect such integrated environments to become vitally important in project-based education.
The situating analytics approach can be expected to, further, add value to writing tools, such as Word and Docs.

A bonus is that the feedback features have the potential to be repurposed in professional design and writing work.
Inasmuch as students learn and create in integrated environments, they are likely to want to keep using them, as they graduate and become professionals. 
Project-oriented work, involving creative tasks, is the least susceptible to being eliminated by automation \cite{national2018workforce}.
Ironically, the incorporation of AI\textemdash as scaffolding in design education\textemdash into project-based work, thus has the potential to become a mainstay of human involvement and performance in the future of work.

Understanding beneficial practices and challenges is critical to developing AI support for educating skilled designers, who can solve complex, sociocultural problems \cite{broadbent2003design}. 
Building a community of stakeholders, through a co-design approach, creates a supportive environment and provides a foundation for sharing complementary expertise.
This is vital for conceptualizing transformational tools, techniques, and environments \cite{fischer2007designing}. 
Future research can beneficially involve more stakeholders\textemdash such as students, teaching assistants, graders, administrators, industry representatives\textemdash and span a range of institutions.
It can employ additional qualitative methods\textemdash such as participant observation and contextual inquiry during various forms of assessments within a course\textemdash to further understand situated practices and identify contextual properties. 
Further, future research can further develop and validate situating analytics for design education.
It can investigate how situating analytics can contribute both to other design contexts and to other educational contexts.

This research has the potential to enable transforming the role of AI in project-based education and work. 
New forms of computation can take support roles, based in situating analytics that provide on demand assessment and feedback.
Situating analytics will integrate interaction with them into user experiences, supporting learning and work.
That is, derivation of and interaction with analytics need to be interwoven, in order for the analytics to be rendered meaningful and actionable components of education and work experiences.
The role of algorithms-in-the-loop is to give assistance.
We do not advocate replacing or reducing instructors. 
Rather, we theorize situating analytics as an alternative, complementary channel or modality of assessment and feedback, which can add value to design education.
In this vision, instructors and students sustain as inceptors, facilitators, and arbiters of creativity.

\bibliographystyle{ACM-Reference-Format}
\bibliography{references}


\begin{thebibliography}{158}


\ifx \showCODEN    \undefined \def \showCODEN     #1{\unskip}     \fi
\ifx \showDOI      \undefined \def \showDOI       #1{#1}\fi
\ifx \showISBNx    \undefined \def \showISBNx     #1{\unskip}     \fi
\ifx \showISBNxiii \undefined \def \showISBNxiii  #1{\unskip}     \fi
\ifx \showISSN     \undefined \def \showISSN      #1{\unskip}     \fi
\ifx \showLCCN     \undefined \def \showLCCN      #1{\unskip}     \fi
\ifx \shownote     \undefined \def \shownote      #1{#1}          \fi
\ifx \showarticletitle \undefined \def \showarticletitle #1{#1}   \fi
\ifx \showURL      \undefined \def \showURL       {\relax}        \fi
\providecommand\bibfield[2]{#2}
\providecommand\bibinfo[2]{#2}
\providecommand\natexlab[1]{#1}
\providecommand\showeprint[2][]{arXiv:#2}

\bibitem[\protect\citeauthoryear{Adams, Forin, Chua, and Radcliffe}{Adams
  et~al\mbox{.}}{2016}]%
        {adams2016characterizing}
\bibfield{author}{\bibinfo{person}{Robin~S Adams}, \bibinfo{person}{Tiago
  Forin}, \bibinfo{person}{Mel Chua}, {and} \bibinfo{person}{David Radcliffe}.}
  \bibinfo{year}{2016}\natexlab{}.
\newblock \showarticletitle{{Characterizing the work of coaching during design
  reviews}}.
\newblock \bibinfo{journal}{\emph{Design Studies}}  \bibinfo{volume}{45}
  (\bibinfo{year}{2016}), \bibinfo{pages}{30--67}.
\newblock


\bibitem[\protect\citeauthoryear{Adobe}{Adobe}{2020}]%
        {adobeSparkTemplate}
\bibfield{author}{\bibinfo{person}{Adobe}.} \bibinfo{year}{2020}\natexlab{}.
\newblock \bibinfo{title}{{Free Poster Template with images from
  https://www.nasa.gov/ and text based on
  thrillist.com/tech/11-nasa-technologies-we-use-every-day | Adobe Spark}}.
\newblock \bibinfo{howpublished}{https://spark.adobe.com/post/1TK56DIfH24Xz/}.
\newblock
\newblock
\shownote{Last accessed: 2020-10-19.}


\bibitem[\protect\citeauthoryear{Arnold and Pistilli}{Arnold and
  Pistilli}{2012}]%
        {arnold2012course}
\bibfield{author}{\bibinfo{person}{Kimberly~E Arnold} {and}
  \bibinfo{person}{Matthew~D Pistilli}.} \bibinfo{year}{2012}\natexlab{}.
\newblock \showarticletitle{{Course signals at Purdue: Using learning analytics
  to increase student success}}. In \bibinfo{booktitle}{\emph{Proceedings of
  the 2nd international conference on learning analytics and knowledge}}. ACM,
  \bibinfo{pages}{267--270}.
\newblock


\bibitem[\protect\citeauthoryear{Ba{\c{c}}{\~a}o, Lobo, and
  Painho}{Ba{\c{c}}{\~a}o et~al\mbox{.}}{2005}]%
        {baccao2005self}
\bibfield{author}{\bibinfo{person}{Fernando Ba{\c{c}}{\~a}o},
  \bibinfo{person}{Victor Lobo}, {and} \bibinfo{person}{Marco Painho}.}
  \bibinfo{year}{2005}\natexlab{}.
\newblock \showarticletitle{The self-organizing map, the Geo-SOM, and relevant
  variants for geosciences}.
\newblock \bibinfo{journal}{\emph{Computers \& Geosciences}}
  \bibinfo{volume}{31}, \bibinfo{number}{2} (\bibinfo{year}{2005}),
  \bibinfo{pages}{155--163}.
\newblock


\bibitem[\protect\citeauthoryear{Baddeley}{Baddeley}{1992}]%
        {baddeley1992working}
\bibfield{author}{\bibinfo{person}{Alan Baddeley}.}
  \bibinfo{year}{1992}\natexlab{}.
\newblock \showarticletitle{Working memory}.
\newblock \bibinfo{journal}{\emph{Science}} \bibinfo{volume}{255},
  \bibinfo{number}{5044} (\bibinfo{year}{1992}), \bibinfo{pages}{556--559}.
\newblock


\bibitem[\protect\citeauthoryear{Barba}{Barba}{2019}]%
        {barba2019cognitive}
\bibfield{author}{\bibinfo{person}{Evan Barba}.}
  \bibinfo{year}{2019}\natexlab{}.
\newblock \showarticletitle{{Cognitive Point of View in Recursive Design}}.
\newblock \bibinfo{journal}{\emph{She Ji: The Journal of Design, Economics, and
  Innovation}} \bibinfo{volume}{5}, \bibinfo{number}{2} (\bibinfo{year}{2019}),
  \bibinfo{pages}{147--162}.
\newblock


\bibitem[\protect\citeauthoryear{Beatty and Ulasewicz}{Beatty and
  Ulasewicz}{2006}]%
        {beatty2006faculty}
\bibfield{author}{\bibinfo{person}{Brian Beatty} {and} \bibinfo{person}{Connie
  Ulasewicz}.} \bibinfo{year}{2006}\natexlab{}.
\newblock \showarticletitle{{Faculty perspectives on moving from Blackboard to
  the Moodle learning management system}}.
\newblock \bibinfo{journal}{\emph{TechTrends}} \bibinfo{volume}{50},
  \bibinfo{number}{4} (\bibinfo{year}{2006}), \bibinfo{pages}{36--45}.
\newblock


\bibitem[\protect\citeauthoryear{Bellamy, Dey, Hind, Hoffman, Houde, Kannan,
  Lohia, Martino, Mehta, Mojsilovic, et~al\mbox{.}}{Bellamy
  et~al\mbox{.}}{2018}]%
        {bellamy2018ai}
\bibfield{author}{\bibinfo{person}{Rachel~KE Bellamy}, \bibinfo{person}{Kuntal
  Dey}, \bibinfo{person}{Michael Hind}, \bibinfo{person}{Samuel~C Hoffman},
  \bibinfo{person}{Stephanie Houde}, \bibinfo{person}{Kalapriya Kannan},
  \bibinfo{person}{Pranay Lohia}, \bibinfo{person}{Jacquelyn Martino},
  \bibinfo{person}{Sameep Mehta}, \bibinfo{person}{Aleksandra Mojsilovic},
  {et~al\mbox{.}}} \bibinfo{year}{2018}\natexlab{}.
\newblock \showarticletitle{AI Fairness 360: An extensible toolkit for
  detecting, understanding, and mitigating unwanted algorithmic bias}.
\newblock \bibinfo{journal}{\emph{arXiv preprint arXiv:1810.01943}}
  (\bibinfo{year}{2018}).
\newblock


\bibitem[\protect\citeauthoryear{Bellotti and Edwards}{Bellotti and
  Edwards}{2001}]%
        {bellotti2001intelligibility}
\bibfield{author}{\bibinfo{person}{Victoria Bellotti} {and}
  \bibinfo{person}{Keith Edwards}.} \bibinfo{year}{2001}\natexlab{}.
\newblock \showarticletitle{Intelligibility and accountability: human
  considerations in context-aware systems}.
\newblock \bibinfo{journal}{\emph{Human--Computer Interaction}}
  \bibinfo{volume}{16}, \bibinfo{number}{2-4} (\bibinfo{year}{2001}),
  \bibinfo{pages}{193--212}.
\newblock


\bibitem[\protect\citeauthoryear{Berners-Lee, Cailliau, Luotonen, Nielsen, and
  Secret}{Berners-Lee et~al\mbox{.}}{1994}]%
        {berners1994world}
\bibfield{author}{\bibinfo{person}{Tim Berners-Lee}, \bibinfo{person}{Robert
  Cailliau}, \bibinfo{person}{Ari Luotonen}, \bibinfo{person}{Henrik~Frystyk
  Nielsen}, {and} \bibinfo{person}{Arthur Secret}.}
  \bibinfo{year}{1994}\natexlab{}.
\newblock \showarticletitle{The world-wide web}.
\newblock \bibinfo{journal}{\emph{Commun. ACM}} \bibinfo{volume}{37},
  \bibinfo{number}{8} (\bibinfo{year}{1994}), \bibinfo{pages}{76--82}.
\newblock


\bibitem[\protect\citeauthoryear{Bertin}{Bertin}{1983}]%
        {bertin1983semiology}
\bibfield{author}{\bibinfo{person}{Jacques Bertin}.}
  \bibinfo{year}{1983}\natexlab{}.
\newblock \bibinfo{booktitle}{\emph{Semiology of Graphics}}.
\newblock \bibinfo{publisher}{University of Wisconsin Press}.
\newblock


\bibitem[\protect\citeauthoryear{Biernacki and Waldorf}{Biernacki and
  Waldorf}{1981}]%
        {biernacki1981snowball}
\bibfield{author}{\bibinfo{person}{Patrick Biernacki} {and}
  \bibinfo{person}{Dan Waldorf}.} \bibinfo{year}{1981}\natexlab{}.
\newblock \showarticletitle{{Snowball sampling: Problems and techniques of
  chain referral sampling}}.
\newblock \bibinfo{journal}{\emph{Sociological methods {\&} research}}
  \bibinfo{volume}{10}, \bibinfo{number}{2} (\bibinfo{year}{1981}),
  \bibinfo{pages}{141--163}.
\newblock


\bibitem[\protect\citeauthoryear{Blikstein}{Blikstein}{2011}]%
        {blikstein2011using}
\bibfield{author}{\bibinfo{person}{Paulo Blikstein}.}
  \bibinfo{year}{2011}\natexlab{}.
\newblock \showarticletitle{{Using learning analytics to assess students'
  behavior in open-ended programming tasks}}. In
  \bibinfo{booktitle}{\emph{Proceedings of the 1st international conference on
  learning analytics and knowledge}}. ACM, \bibinfo{pages}{110--116}.
\newblock


\bibitem[\protect\citeauthoryear{Brandt, Cennamo, Douglas, Vernon, McGrath, and
  Reimer}{Brandt et~al\mbox{.}}{2013}]%
        {brandt2013theoretical}
\bibfield{author}{\bibinfo{person}{Carol~B Brandt}, \bibinfo{person}{Katherine
  Cennamo}, \bibinfo{person}{Sarah Douglas}, \bibinfo{person}{Mitzi Vernon},
  \bibinfo{person}{Margarita McGrath}, {and} \bibinfo{person}{Yolanda Reimer}.}
  \bibinfo{year}{2013}\natexlab{}.
\newblock \showarticletitle{{A theoretical framework for the studio as a
  learning environment}}.
\newblock \bibinfo{journal}{\emph{International Journal of Technology and
  Design Education}} \bibinfo{volume}{23}, \bibinfo{number}{2}
  (\bibinfo{year}{2013}), \bibinfo{pages}{329--348}.
\newblock


\bibitem[\protect\citeauthoryear{Britain, Jain, Lupfer, Kerne, Perrine, Seo,
  and Sungkajun}{Britain et~al\mbox{.}}{2020}]%
        {britain2020}
\bibfield{author}{\bibinfo{person}{Gabriel Britain}, \bibinfo{person}{Ajit
  Jain}, \bibinfo{person}{Nic Lupfer}, \bibinfo{person}{Andruid Kerne},
  \bibinfo{person}{Aaron Perrine}, \bibinfo{person}{Jinsil Seo}, {and}
  \bibinfo{person}{Annie Sungkajun}.} \bibinfo{year}{2020}\natexlab{}.
\newblock \showarticletitle{{Design is (A)live: An Environment Integrating
  Ideation and Assessment}}. In \bibinfo{booktitle}{\emph{CHI Late-Breaking
  Work}}. ACM, \bibinfo{pages}{1--8}.
\newblock


\bibitem[\protect\citeauthoryear{Broadbent and Cross}{Broadbent and
  Cross}{2003}]%
        {broadbent2003design}
\bibfield{author}{\bibinfo{person}{John~A Broadbent} {and}
  \bibinfo{person}{Nigel Cross}.} \bibinfo{year}{2003}\natexlab{}.
\newblock \showarticletitle{{Design education in the information age}}.
\newblock \bibinfo{journal}{\emph{Journal of Engineering Design}}
  \bibinfo{volume}{14}, \bibinfo{number}{4} (\bibinfo{year}{2003}),
  \bibinfo{pages}{439--446}.
\newblock


\bibitem[\protect\citeauthoryear{Brusasco, Caneparo, Carrara, Fioravanti,
  Novembri, and Zorgno}{Brusasco et~al\mbox{.}}{2000}]%
        {brusasco2000computer}
\bibfield{author}{\bibinfo{person}{Pio~Luigi Brusasco}, \bibinfo{person}{Luca
  Caneparo}, \bibinfo{person}{Gianfranco Carrara}, \bibinfo{person}{Antonio
  Fioravanti}, \bibinfo{person}{Gabriele Novembri}, {and}
  \bibinfo{person}{Anna~Maria Zorgno}.} \bibinfo{year}{2000}\natexlab{}.
\newblock \showarticletitle{{Computer supported design studio}}.
\newblock \bibinfo{journal}{\emph{Automation in Construction}}
  \bibinfo{volume}{9}, \bibinfo{number}{4} (\bibinfo{year}{2000}),
  \bibinfo{pages}{393--408}.
\newblock


\bibitem[\protect\citeauthoryear{Buchanan}{Buchanan}{1992}]%
        {buchanan1992wicked}
\bibfield{author}{\bibinfo{person}{Richard Buchanan}.}
  \bibinfo{year}{1992}\natexlab{}.
\newblock \showarticletitle{{Wicked problems in design thinking}}.
\newblock \bibinfo{journal}{\emph{Design issues}} \bibinfo{volume}{8},
  \bibinfo{number}{2} (\bibinfo{year}{1992}), \bibinfo{pages}{5--21}.
\newblock


\bibitem[\protect\citeauthoryear{Card, Mackinlay, and Shneiderman}{Card
  et~al\mbox{.}}{1999}]%
        {card1999readings}
\bibfield{author}{\bibinfo{person}{Stuart Card}, \bibinfo{person}{JD
  Mackinlay}, {and} \bibinfo{person}{B Shneiderman}.}
  \bibinfo{year}{1999}\natexlab{}.
\newblock \bibinfo{booktitle}{\emph{{Readings in information visualization:
  using vision to think}}}.
\newblock \bibinfo{publisher}{Morgan Kaufmann}.
\newblock


\bibitem[\protect\citeauthoryear{Charmaz}{Charmaz}{2014}]%
        {charmaz2014constructing}
\bibfield{author}{\bibinfo{person}{Kathy Charmaz}.}
  \bibinfo{year}{2014}\natexlab{}.
\newblock \bibinfo{booktitle}{\emph{{Constructing grounded theory}}}.
\newblock \bibinfo{publisher}{Sage}.
\newblock


\bibitem[\protect\citeauthoryear{Chen}{Chen}{2016}]%
        {chen2016exploring}
\bibfield{author}{\bibinfo{person}{Wenzhi Chen}.}
  \bibinfo{year}{2016}\natexlab{}.
\newblock \showarticletitle{{Exploring the learning problems and resource usage
  of undergraduate industrial design students in design studio courses}}.
\newblock \bibinfo{journal}{\emph{International Journal of Technology and
  Design Education}} \bibinfo{volume}{26}, \bibinfo{number}{3}
  (\bibinfo{year}{2016}), \bibinfo{pages}{461--487}.
\newblock


\bibitem[\protect\citeauthoryear{Crismond and Adams}{Crismond and
  Adams}{2012}]%
        {crismond2012informed}
\bibfield{author}{\bibinfo{person}{David~P Crismond} {and}
  \bibinfo{person}{Robin~S Adams}.} \bibinfo{year}{2012}\natexlab{}.
\newblock \showarticletitle{{The informed design teaching and learning
  matrix}}.
\newblock \bibinfo{journal}{\emph{Journal of Engineering Education}}
  \bibinfo{volume}{101}, \bibinfo{number}{4} (\bibinfo{year}{2012}),
  \bibinfo{pages}{738--797}.
\newblock


\bibitem[\protect\citeauthoryear{Cross}{Cross}{1982}]%
        {cross1982designerly}
\bibfield{author}{\bibinfo{person}{Nigel Cross}.}
  \bibinfo{year}{1982}\natexlab{}.
\newblock \showarticletitle{{Designerly ways of knowing}}.
\newblock \bibinfo{journal}{\emph{Design studies}} \bibinfo{volume}{3},
  \bibinfo{number}{4} (\bibinfo{year}{1982}), \bibinfo{pages}{221--227}.
\newblock


\bibitem[\protect\citeauthoryear{Dannels, Gaffney, and Martin}{Dannels
  et~al\mbox{.}}{2008}]%
        {dannels2008beyond}
\bibfield{author}{\bibinfo{person}{Deanna Dannels},
  \bibinfo{person}{Amy~Housley Gaffney}, {and} \bibinfo{person}{Kelly~Norris
  Martin}.} \bibinfo{year}{2008}\natexlab{}.
\newblock \showarticletitle{{Beyond content, deeper than delivery: What
  critique feedback reveals about communication expectations in design
  education}}.
\newblock \bibinfo{journal}{\emph{International Journal for the Scholarship of
  teaching and Learning}} \bibinfo{volume}{2}, \bibinfo{number}{2}
  (\bibinfo{year}{2008}), \bibinfo{pages}{12}.
\newblock


\bibitem[\protect\citeauthoryear{Dannels, {Housley Gaffney}, and
  Martin}{Dannels et~al\mbox{.}}{2011}]%
        {dannels2011students}
\bibfield{author}{\bibinfo{person}{Deanna~P Dannels}, \bibinfo{person}{Amy~L
  {Housley Gaffney}}, {and} \bibinfo{person}{Kelly~Norris Martin}.}
  \bibinfo{year}{2011}\natexlab{}.
\newblock \showarticletitle{{Students' talk about the climate of feedback
  interventions in the critique}}.
\newblock \bibinfo{journal}{\emph{Communication Education}}
  \bibinfo{volume}{60}, \bibinfo{number}{1} (\bibinfo{year}{2011}),
  \bibinfo{pages}{95--114}.
\newblock


\bibitem[\protect\citeauthoryear{Davis}{Davis}{2017}]%
        {davis2017teaching}
\bibfield{author}{\bibinfo{person}{Meredith Davis}.}
  \bibinfo{year}{2017}\natexlab{}.
\newblock \bibinfo{booktitle}{\emph{{Teaching Design: a guide to curriculum and
  pedagogy for college design faculty and teachers who use design in their
  classrooms}}}.
\newblock \bibinfo{publisher}{Simon and Schuster}.
\newblock


\bibitem[\protect\citeauthoryear{Dawson, Macfadyen, Evan, Foulsham, and
  Kingstone}{Dawson et~al\mbox{.}}{2012}]%
        {dawson2012using}
\bibfield{author}{\bibinfo{person}{Shane Dawson}, \bibinfo{person}{Leah
  Macfadyen}, \bibinfo{person}{F~Risko Evan}, \bibinfo{person}{Tom Foulsham},
  {and} \bibinfo{person}{Alan Kingstone}.} \bibinfo{year}{2012}\natexlab{}.
\newblock \showarticletitle{{Using technology to encourage self-directed
  learning: The Collaborative Lecture Annotation System (CLAS)}}. In
  \bibinfo{booktitle}{\emph{Australasian Society for Computers in Learning in
  Tertiatry Education}}. \bibinfo{pages}{246--255}.
\newblock


\bibitem[\protect\citeauthoryear{{De La Harpe}, Peterson, Frankham, Zehner,
  Neale, Musgrave, and McDermott}{{De La Harpe} et~al\mbox{.}}{2009}]%
        {de2009assessment}
\bibfield{author}{\bibinfo{person}{Barbara {De La Harpe}},
  \bibinfo{person}{J~Fiona Peterson}, \bibinfo{person}{Noel Frankham},
  \bibinfo{person}{Robert Zehner}, \bibinfo{person}{Douglas Neale},
  \bibinfo{person}{Elizabeth Musgrave}, {and} \bibinfo{person}{Ruth
  McDermott}.} \bibinfo{year}{2009}\natexlab{}.
\newblock \showarticletitle{{Assessment focus in studio: What is most prominent
  in architecture, art and design?}}
\newblock \bibinfo{journal}{\emph{International Journal of Art {\&} Design
  Education}} \bibinfo{volume}{28}, \bibinfo{number}{1} (\bibinfo{year}{2009}),
  \bibinfo{pages}{37--51}.
\newblock


\bibitem[\protect\citeauthoryear{DiSalvo and DiSalvo}{DiSalvo and
  DiSalvo}{2014}]%
        {disalvo2014designing}
\bibfield{author}{\bibinfo{person}{Betsy DiSalvo} {and} \bibinfo{person}{Carl
  DiSalvo}.} \bibinfo{year}{2014}\natexlab{}.
\newblock \showarticletitle{{Designing for democracy in education:
  Participatory design and the learning sciences}}.
\newblock In \bibinfo{booktitle}{\emph{Proceedings of the Eleventh
  International Conference of the Learning Sciences}}.
  \bibinfo{publisher}{Boulder, CO: International Society of the Learning
  Sciences}, \bibinfo{pages}{793--799}.
\newblock


\bibitem[\protect\citeauthoryear{Dourish}{Dourish}{2004}]%
        {dourish2004we}
\bibfield{author}{\bibinfo{person}{Paul Dourish}.}
  \bibinfo{year}{2004}\natexlab{}.
\newblock \showarticletitle{What we talk about when we talk about context}.
\newblock \bibinfo{journal}{\emph{Personal and ubiquitous computing}}
  \bibinfo{volume}{8}, \bibinfo{number}{1} (\bibinfo{year}{2004}),
  \bibinfo{pages}{19--30}.
\newblock


\bibitem[\protect\citeauthoryear{Dove, Halskov, Forlizzi, and Zimmerman}{Dove
  et~al\mbox{.}}{2017}]%
        {dove2017ux}
\bibfield{author}{\bibinfo{person}{Graham Dove}, \bibinfo{person}{Kim Halskov},
  \bibinfo{person}{Jodi Forlizzi}, {and} \bibinfo{person}{John Zimmerman}.}
  \bibinfo{year}{2017}\natexlab{}.
\newblock \showarticletitle{UX design innovation: Challenges for working with
  machine learning as a design material}. In
  \bibinfo{booktitle}{\emph{Proceedings of the 2017 chi conference on human
  factors in computing systems}}. \bibinfo{pages}{278--288}.
\newblock


\bibitem[\protect\citeauthoryear{Dow, Gerber, and Wong}{Dow
  et~al\mbox{.}}{2013}]%
        {dow2013pilot}
\bibfield{author}{\bibinfo{person}{Steven Dow}, \bibinfo{person}{Elizabeth
  Gerber}, {and} \bibinfo{person}{Audris Wong}.}
  \bibinfo{year}{2013}\natexlab{}.
\newblock \showarticletitle{{A pilot study of using crowds in the classroom}}.
  In \bibinfo{booktitle}{\emph{Proceedings of the SIGCHI Conference on Human
  Factors in Computing Systems}}. ACM, \bibinfo{pages}{227--236}.
\newblock


\bibitem[\protect\citeauthoryear{Dutton}{Dutton}{1987}]%
        {dutton1987design}
\bibfield{author}{\bibinfo{person}{Thomas~A Dutton}.}
  \bibinfo{year}{1987}\natexlab{}.
\newblock \showarticletitle{{Design and studio pedagogy}}.
\newblock \bibinfo{journal}{\emph{Journal of architectural education}}
  \bibinfo{volume}{41}, \bibinfo{number}{1} (\bibinfo{year}{1987}),
  \bibinfo{pages}{16--25}.
\newblock


\bibitem[\protect\citeauthoryear{Duval}{Duval}{2011}]%
        {duval2011attention}
\bibfield{author}{\bibinfo{person}{Erik Duval}.}
  \bibinfo{year}{2011}\natexlab{}.
\newblock \showarticletitle{{Attention please!: learning analytics for
  visualization and recommendation}}. In \bibinfo{booktitle}{\emph{Proceedings
  of the 1st international conference on learning analytics and knowledge}}.
  ACM, \bibinfo{pages}{9--17}.
\newblock


\bibitem[\protect\citeauthoryear{Erickson}{Erickson}{2002}]%
        {erickson2002concept}
\bibfield{author}{\bibinfo{person}{H~Lynn Erickson}.}
  \bibinfo{year}{2002}\natexlab{}.
\newblock \bibinfo{booktitle}{\emph{Concept-based curriculum and instruction:
  Teaching beyond the facts}}.
\newblock \bibinfo{publisher}{Corwin Press}.
\newblock


\bibitem[\protect\citeauthoryear{Estivill-Castro and Lee}{Estivill-Castro and
  Lee}{2002}]%
        {estivill2002multi}
\bibfield{author}{\bibinfo{person}{Vladimir Estivill-Castro} {and}
  \bibinfo{person}{Ickjai Lee}.} \bibinfo{year}{2002}\natexlab{}.
\newblock \showarticletitle{Multi-level clustering and its visualization for
  exploratory spatial analysis}.
\newblock \bibinfo{journal}{\emph{GeoInformatica}} \bibinfo{volume}{6},
  \bibinfo{number}{2} (\bibinfo{year}{2002}), \bibinfo{pages}{123--152}.
\newblock


\bibitem[\protect\citeauthoryear{Few}{Few}{2013}]%
        {few2013information}
\bibfield{author}{\bibinfo{person}{Stephen Few}.}
  \bibinfo{year}{2013}\natexlab{}.
\newblock \bibinfo{booktitle}{\emph{Information Dashboard Design: Displaying
  data for at-a-glance monitoring}}. Vol.~\bibinfo{volume}{5}.
\newblock \bibinfo{publisher}{Analytics Press Burlingame, CA}.
\newblock


\bibitem[\protect\citeauthoryear{Fischer}{Fischer}{2007}]%
        {fischer2007designing}
\bibfield{author}{\bibinfo{person}{Gerhard Fischer}.}
  \bibinfo{year}{2007}\natexlab{}.
\newblock \showarticletitle{{Designing socio-technical environments in support
  of meta-design and social creativity}}. In
  \bibinfo{booktitle}{\emph{Proceedings of the 8th iternational conference on
  Computer supported collaborative learning}}. International Society of the
  Learning Sciences, \bibinfo{pages}{2--11}.
\newblock


\bibitem[\protect\citeauthoryear{for Everything}{for Everything}{2019}]%
        {noun_project_icons}
\bibfield{author}{\bibinfo{person}{Noun Project: Free~Icons for Everything}.}
  \bibinfo{year}{2019}\natexlab{}.
\newblock \bibinfo{title}{{Dashboard by Rafael Garcia Motta from the Noun
  Project; dashboard by Aybige from the Noun Project; students by Philipp
  Petzka from the Noun Project; and workshop by DailyPM from the Noun
  Project.}}
\newblock
\newblock
\urldef\tempurl%
\url{https://thenounproject.com/}
\showURL{%
Retrieved 2019-11-26 from \tempurl}


\bibitem[\protect\citeauthoryear{Frauenberger, Good, and
  Keay-Bright}{Frauenberger et~al\mbox{.}}{2011}]%
        {frauenberger2011designing}
\bibfield{author}{\bibinfo{person}{Christopher Frauenberger},
  \bibinfo{person}{Judith Good}, {and} \bibinfo{person}{Wendy Keay-Bright}.}
  \bibinfo{year}{2011}\natexlab{}.
\newblock \showarticletitle{Designing technology for children with special
  needs: bridging perspectives through participatory design}.
\newblock \bibinfo{journal}{\emph{CoDesign}} \bibinfo{volume}{7},
  \bibinfo{number}{1} (\bibinfo{year}{2011}), \bibinfo{pages}{1--28}.
\newblock


\bibitem[\protect\citeauthoryear{Frich, {Mose Biskjaer}, and Dalsgaard}{Frich
  et~al\mbox{.}}{2018}]%
        {frich2018twenty}
\bibfield{author}{\bibinfo{person}{Jonas Frich}, \bibinfo{person}{Michael {Mose
  Biskjaer}}, {and} \bibinfo{person}{Peter Dalsgaard}.}
  \bibinfo{year}{2018}\natexlab{}.
\newblock \showarticletitle{{Twenty Years of Creativity Research in
  Human-Computer Interaction: Current State and Future Directions}}. In
  \bibinfo{booktitle}{\emph{Proceedings of the 2018 Designing Interactive
  Systems Conference}}. ACM, \bibinfo{pages}{1235--1257}.
\newblock


\bibitem[\protect\citeauthoryear{Fu, Chan, Cagan, Kotovsky, Schunn, and
  Wood}{Fu et~al\mbox{.}}{2013}]%
        {fu2013meaning}
\bibfield{author}{\bibinfo{person}{Katherine Fu}, \bibinfo{person}{Joel Chan},
  \bibinfo{person}{Jonathan Cagan}, \bibinfo{person}{Kenneth Kotovsky},
  \bibinfo{person}{Christian Schunn}, {and} \bibinfo{person}{Kristin Wood}.}
  \bibinfo{year}{2013}\natexlab{}.
\newblock \showarticletitle{The meaning of “near” and “far”: the impact
  of structuring design databases and the effect of distance of analogy on
  design output}.
\newblock \bibinfo{journal}{\emph{Journal of Mechanical Design}}
  \bibinfo{volume}{135}, \bibinfo{number}{2} (\bibinfo{year}{2013}).
\newblock


\bibitem[\protect\citeauthoryear{Gentner}{Gentner}{1983}]%
        {gentner1983structure}
\bibfield{author}{\bibinfo{person}{Dedre Gentner}.}
  \bibinfo{year}{1983}\natexlab{}.
\newblock \showarticletitle{{Structure-mapping: A theoretical framework for
  analogy}}.
\newblock \bibinfo{journal}{\emph{Cognitive science}} \bibinfo{volume}{7},
  \bibinfo{number}{2} (\bibinfo{year}{1983}), \bibinfo{pages}{155--170}.
\newblock


\bibitem[\protect\citeauthoryear{Gero and Maher}{Gero and Maher}{1991}]%
        {gero1991mutation}
\bibfield{author}{\bibinfo{person}{John~S Gero} {and} \bibinfo{person}{Mary~Lou
  Maher}.} \bibinfo{year}{1991}\natexlab{}.
\newblock \showarticletitle{Mutation and analogy to support creativity in
  computer-aided design}.
\newblock \bibinfo{journal}{\emph{CAAD Futures'91}} (\bibinfo{year}{1991}),
  \bibinfo{pages}{241--249}.
\newblock


\bibitem[\protect\citeauthoryear{Giaccardi and Karana}{Giaccardi and
  Karana}{2015}]%
        {giaccardi2015MaterialsExperience}
\bibfield{author}{\bibinfo{person}{Elisa Giaccardi} {and}
  \bibinfo{person}{Elvin Karana}.} \bibinfo{year}{2015}\natexlab{}.
\newblock \showarticletitle{Foundations of Materials Experience: An Approach
  for HCI}. In \bibinfo{booktitle}{\emph{Proceedings of the 33rd Annual ACM
  Conference on Human Factors in Computing Systems}} (Seoul, Republic of Korea)
  \emph{(\bibinfo{series}{CHI '15})}. \bibinfo{publisher}{Association for
  Computing Machinery}, \bibinfo{address}{New York, NY, USA},
  \bibinfo{pages}{2447--2456}.
\newblock
\showISBNx{9781450331456}
\urldef\tempurl%
\url{https://doi.org/10.1145/2702123.2702337}
\showDOI{\tempurl}


\bibitem[\protect\citeauthoryear{Glenberg and Langston}{Glenberg and
  Langston}{1992}]%
        {glenberg1992comprehension}
\bibfield{author}{\bibinfo{person}{Arthur~M Glenberg} {and}
  \bibinfo{person}{William~E Langston}.} \bibinfo{year}{1992}\natexlab{}.
\newblock \showarticletitle{Comprehension of illustrated text: Pictures help to
  build mental models}.
\newblock \bibinfo{journal}{\emph{Journal of memory and language}}
  \bibinfo{volume}{31}, \bibinfo{number}{2} (\bibinfo{year}{1992}),
  \bibinfo{pages}{129--151}.
\newblock


\bibitem[\protect\citeauthoryear{Gray}{Gray}{2018}]%
        {gray2018democratizing}
\bibfield{author}{\bibinfo{person}{Colin~M Gray}.}
  \bibinfo{year}{2018}\natexlab{}.
\newblock \showarticletitle{{Democratizing assessment practices through
  multimodal critique in the design classroom}}.
\newblock \bibinfo{journal}{\emph{International Journal of Technology and
  Design Education}} (\bibinfo{year}{2018}), \bibinfo{pages}{1--18}.
\newblock


\bibitem[\protect\citeauthoryear{Gray, Exter, and Krause}{Gray
  et~al\mbox{.}}{2017}]%
        {gray2017individual}
\bibfield{author}{\bibinfo{person}{Colin~M Gray}, \bibinfo{person}{Marisa
  Exter}, {and} \bibinfo{person}{Terri~S Krause}.}
  \bibinfo{year}{2017}\natexlab{}.
\newblock \showarticletitle{{Moving Towards Individual Competence From Group
  Work in Transdisciplinary Education}}. In \bibinfo{booktitle}{\emph{2017 ASEE
  Annual Conference {\&} Exposition}}. \bibinfo{publisher}{ASEE Conferences},
  \bibinfo{address}{Columbus, Ohio}.
\newblock
\newblock
\shownote{https://peer.asee.org/28691.}


\bibitem[\protect\citeauthoryear{Green and Chen}{Green and Chen}{2019a}]%
        {green2019disparate}
\bibfield{author}{\bibinfo{person}{Ben Green} {and} \bibinfo{person}{Yiling
  Chen}.} \bibinfo{year}{2019}\natexlab{a}.
\newblock \showarticletitle{{Disparate interactions: An algorithm-in-the-loop
  analysis of fairness in risk assessments}}. In
  \bibinfo{booktitle}{\emph{Proceedings of the Conference on Fairness,
  Accountability, and Transparency}}. ACM, \bibinfo{pages}{90--99}.
\newblock


\bibitem[\protect\citeauthoryear{Green and Chen}{Green and Chen}{2019b}]%
        {green2019principles}
\bibfield{author}{\bibinfo{person}{Ben Green} {and} \bibinfo{person}{Yiling
  Chen}.} \bibinfo{year}{2019}\natexlab{b}.
\newblock \showarticletitle{{The principles and limits of algorithm-in-the-loop
  decision making}}.
\newblock \bibinfo{journal}{\emph{Proceedings of the ACM on Human-Computer
  Interaction}} \bibinfo{volume}{3}, \bibinfo{number}{CSCW}
  (\bibinfo{year}{2019}), \bibinfo{pages}{50}.
\newblock


\bibitem[\protect\citeauthoryear{Greene, Freed, and Sawyer}{Greene
  et~al\mbox{.}}{2019}]%
        {greene2019fostering}
\bibfield{author}{\bibinfo{person}{Jeffrey~A Greene}, \bibinfo{person}{Rebekah
  Freed}, {and} \bibinfo{person}{R~Keith Sawyer}.}
  \bibinfo{year}{2019}\natexlab{}.
\newblock \showarticletitle{{Fostering creative performance in art and design
  education via self-regulated learning}}.
\newblock \bibinfo{journal}{\emph{Instructional Science}} \bibinfo{volume}{47},
  \bibinfo{number}{2} (\bibinfo{year}{2019}), \bibinfo{pages}{127--149}.
\newblock


\bibitem[\protect\citeauthoryear{Guilford}{Guilford}{1950}]%
        {guilford1950}
\bibfield{author}{\bibinfo{person}{J.P. Guilford}.}
  \bibinfo{year}{1950}\natexlab{}.
\newblock \showarticletitle{Creativity}.
\newblock \bibinfo{journal}{\emph{American Psychologist}}  \bibinfo{volume}{5}
  (\bibinfo{year}{1950}), \bibinfo{pages}{444--454}.
\newblock


\bibitem[\protect\citeauthoryear{Hamilton, Lupfer, Botello, Tesch, Stacy,
  Merrill, Williford, Bentley, and Kerne}{Hamilton et~al\mbox{.}}{2018}]%
        {hamilton2018collaborative}
\bibfield{author}{\bibinfo{person}{William~A Hamilton}, \bibinfo{person}{Nic
  Lupfer}, \bibinfo{person}{Nicolas Botello}, \bibinfo{person}{Tyler Tesch},
  \bibinfo{person}{Alex Stacy}, \bibinfo{person}{Jeremy Merrill},
  \bibinfo{person}{Blake Williford}, \bibinfo{person}{Frank~R Bentley}, {and}
  \bibinfo{person}{Andruid Kerne}.} \bibinfo{year}{2018}\natexlab{}.
\newblock \showarticletitle{{Collaborative Live Media Curation: Shared Context
  for Participation in Online Learning}}. In
  \bibinfo{booktitle}{\emph{Proceedings of the 2018 CHI Conference on Human
  Factors in Computing Systems}}. ACM, \bibinfo{pages}{1--14}.
\newblock


\bibitem[\protect\citeauthoryear{Higgins}{Higgins}{2002}]%
        {higgins2002fluxus}
\bibfield{author}{\bibinfo{person}{Hannah Higgins}.}
  \bibinfo{year}{2002}\natexlab{}.
\newblock \bibinfo{booktitle}{\emph{{Fluxus experience}}}.
\newblock \bibinfo{publisher}{Univ of California Press}.
\newblock


\bibitem[\protect\citeauthoryear{Hinton}{Hinton}{2020}]%
        {hinton2020tweet}
\bibfield{author}{\bibinfo{person}{Geoffrey Hinton}.}
  \bibinfo{year}{2020}\natexlab{}.
\newblock \bibinfo{title}{{Geoffrey Hinton on Twitter: "Suppose you have cancer
  and you have to choose between a black box AI surgeon that cannot explain how
  it works but has a 90{\%} cure rate and a human surgeon with an 80{\%} cure
  rate. Do you want the AI surgeon to be illegal?"}}.
\newblock
\newblock
\urldef\tempurl%
\url{https://twitter.com/geoffreyhinton/status/1230592238490615816}
\showURL{%
Retrieved 2020-03-19 from \tempurl}


\bibitem[\protect\citeauthoryear{How, Cheah, Chan, Khor, and Say}{How
  et~al\mbox{.}}{2020}]%
        {how2020artificial}
\bibfield{author}{\bibinfo{person}{Meng-Leong How}, \bibinfo{person}{Sin-Mei
  Cheah}, \bibinfo{person}{Yong-Jiet Chan}, \bibinfo{person}{Aik~Cheow Khor},
  {and} \bibinfo{person}{Eunice Mei~Ping Say}.}
  \bibinfo{year}{2020}\natexlab{}.
\newblock \showarticletitle{Artificial intelligence-enhanced decision support
  for informing global sustainable development: A human-centric AI-thinking
  approach}.
\newblock \bibinfo{journal}{\emph{Information}} \bibinfo{volume}{11},
  \bibinfo{number}{1} (\bibinfo{year}{2020}), \bibinfo{pages}{39}.
\newblock


\bibitem[\protect\citeauthoryear{Hudson and Manning}{Hudson and
  Manning}{2018}]%
        {hudson2018compositional}
\bibfield{author}{\bibinfo{person}{Drew~A Hudson} {and}
  \bibinfo{person}{Christopher~D Manning}.} \bibinfo{year}{2018}\natexlab{}.
\newblock \showarticletitle{{Compositional attention networks for machine
  reasoning}}.
\newblock \bibinfo{journal}{\emph{arXiv preprint arXiv:1803.03067}}
  (\bibinfo{year}{2018}).
\newblock


\bibitem[\protect\citeauthoryear{Hui, Gerber, and Dow}{Hui
  et~al\mbox{.}}{2014}]%
        {hui2014crowd}
\bibfield{author}{\bibinfo{person}{Julie~S Hui}, \bibinfo{person}{Elizabeth~M
  Gerber}, {and} \bibinfo{person}{Steven~P Dow}.}
  \bibinfo{year}{2014}\natexlab{}.
\newblock \showarticletitle{{Crowd-based design activities: helping students
  connect with users online}}. In \bibinfo{booktitle}{\emph{Proceedings of the
  2014 conference on Designing interactive systems}}. ACM,
  \bibinfo{pages}{875--884}.
\newblock


\bibitem[\protect\citeauthoryear{Itten}{Itten}{1970}]%
        {itten1970elements}
\bibfield{author}{\bibinfo{person}{Johannes Itten}.}
  \bibinfo{year}{1970}\natexlab{}.
\newblock \bibinfo{booktitle}{\emph{{The elements of color}}}.
\newblock \bibinfo{publisher}{John Wiley {\&} Sons}.
\newblock


\bibitem[\protect\citeauthoryear{Jain}{Jain}{2017}]%
        {jain2017measuring}
\bibfield{author}{\bibinfo{person}{Ajit Jain}.}
  \bibinfo{year}{2017}\natexlab{}.
\newblock \showarticletitle{Measuring Creativity: Multi-Scale Visual and
  Conceptual Design Analysis}. In \bibinfo{booktitle}{\emph{Proceedings of the
  2017 ACM SIGCHI Conference on Creativity and Cognition}}.
  \bibinfo{pages}{490--495}.
\newblock


\bibitem[\protect\citeauthoryear{Jain}{Jain}{2021}]%
        {jain2021support}
\bibfield{author}{\bibinfo{person}{Ajit Jain}.}
  \bibinfo{year}{2021}\natexlab{}.
\newblock \emph{\bibinfo{title}{How to Support Situated Design Education
  through AI-Based Analytics}}.
\newblock \bibinfo{thesistype}{Ph.D. Dissertation}.
\newblock


\bibitem[\protect\citeauthoryear{Jain, Kasiviswanathan, and Huang}{Jain
  et~al\mbox{.}}{2016}]%
        {jain2016towards}
\bibfield{author}{\bibinfo{person}{Ajit Jain}, \bibinfo{person}{Girish
  Kasiviswanathan}, {and} \bibinfo{person}{Ruihong Huang}.}
  \bibinfo{year}{2016}\natexlab{}.
\newblock \showarticletitle{Towards accurate event detection in social media: A
  weakly supervised approach for learning implicit event indicators}. In
  \bibinfo{booktitle}{\emph{Proceedings of the 2nd Workshop on Noisy
  User-generated Text (WNUT)}}. \bibinfo{pages}{70--77}.
\newblock


\bibitem[\protect\citeauthoryear{Jain, Kerne, Lupfer, Britain, Perrine, Choe,
  Keyser, and Huang}{Jain et~al\mbox{.}}{2021}]%
        {jain2021recognizing}
\bibfield{author}{\bibinfo{person}{Ajit Jain}, \bibinfo{person}{Andruid Kerne},
  \bibinfo{person}{Nic Lupfer}, \bibinfo{person}{Gabriel Britain},
  \bibinfo{person}{Aaron Perrine}, \bibinfo{person}{Yoonsuck Choe},
  \bibinfo{person}{John Keyser}, {and} \bibinfo{person}{Ruihong Huang}.}
  \bibinfo{year}{2021}\natexlab{}.
\newblock \showarticletitle{Recognizing creative visual design: multiscale
  design characteristics in free-form web curation documents}. In
  \bibinfo{booktitle}{\emph{Proceedings of the 21st ACM Symposium on Document
  Engineering}}. \bibinfo{pages}{1--10}.
\newblock


\bibitem[\protect\citeauthoryear{Jain, Kerne, Lupfer, Britain, Perrine, Choe,
  Keyser, Huang, Seo, Sungkajun, et~al\mbox{.}}{Jain et~al\mbox{.}}{2024}]%
        {jain2024indexing}
\bibfield{author}{\bibinfo{person}{Ajit Jain}, \bibinfo{person}{Andruid Kerne},
  \bibinfo{person}{Nic Lupfer}, \bibinfo{person}{Gabriel Britain},
  \bibinfo{person}{Aaron Perrine}, \bibinfo{person}{Yoonsuck Choe},
  \bibinfo{person}{John Keyser}, \bibinfo{person}{Ruihong Huang},
  \bibinfo{person}{Jinsil Seo}, \bibinfo{person}{Annie Sungkajun},
  {et~al\mbox{.}}} \bibinfo{year}{2024}\natexlab{}.
\newblock \showarticletitle{Indexing Analytics to Instances: How Integrating a
  Dashboard can Support Design Education}.
\newblock \bibinfo{journal}{\emph{arXiv preprint arXiv:2404.05417}}
  (\bibinfo{year}{2024}).
\newblock


\bibitem[\protect\citeauthoryear{Jain, Lupfer, Qu, Linder, Kerne, and
  Smith}{Jain et~al\mbox{.}}{2015}]%
        {jain2015tweetbubble}
\bibfield{author}{\bibinfo{person}{Ajit Jain}, \bibinfo{person}{Nic Lupfer},
  \bibinfo{person}{Yin Qu}, \bibinfo{person}{Rhema Linder},
  \bibinfo{person}{Andruid Kerne}, {and} \bibinfo{person}{Steven~M. Smith}.}
  \bibinfo{year}{2015}\natexlab{}.
\newblock \showarticletitle{Evaluating tweetbubble with ideation metrics of
  exploratory browsing}. In \bibinfo{booktitle}{\emph{Proceedings of the 2015
  ACM SIGCHI Conference on Creativity and Cognition}}. \bibinfo{pages}{53--62}.
\newblock


\bibitem[\protect\citeauthoryear{Jiang and Canny}{Jiang and Canny}{2017}]%
        {jiang2017interactive}
\bibfield{author}{\bibinfo{person}{Biye Jiang} {and} \bibinfo{person}{John
  Canny}.} \bibinfo{year}{2017}\natexlab{}.
\newblock \showarticletitle{{Interactive machine learning via a gpu-accelerated
  toolkit}}. In \bibinfo{booktitle}{\emph{Proceedings of the 22nd International
  Conference on Intelligent User Interfaces}}. ACM, \bibinfo{pages}{535--546}.
\newblock


\bibitem[\protect\citeauthoryear{Jung and Do}{Jung and Do}{2000}]%
        {jung2000immersive}
\bibfield{author}{\bibinfo{person}{Thomas Jung} {and} \bibinfo{person}{Ellen
  Yi-Luen Do}.} \bibinfo{year}{2000}\natexlab{}.
\newblock \showarticletitle{{Immersive redliner: collaborative design in
  cyberspace}}. In \bibinfo{booktitle}{\emph{ACADIA Eternity, infinity and
  virtuality in architecture}}. \bibinfo{publisher}{CUMINCAD},
  \bibinfo{pages}{185--194}.
\newblock


\bibitem[\protect\citeauthoryear{Jung, Do, and Gross}{Jung
  et~al\mbox{.}}{1999}]%
        {jung1999immersive}
\bibfield{author}{\bibinfo{person}{Thomas Jung}, \bibinfo{person}{Ellen Yi-Luen
  Do}, {and} \bibinfo{person}{Mark~D Gross}.} \bibinfo{year}{1999}\natexlab{}.
\newblock \showarticletitle{{Immersive redlining and annotation of 3D design
  models on the web}}.
\newblock In \bibinfo{booktitle}{\emph{Computers in Building}}.
  \bibinfo{publisher}{Springer}, \bibinfo{pages}{81--98}.
\newblock


\bibitem[\protect\citeauthoryear{Kaprow}{Kaprow}{2014}]%
        {kaprow2014happenings}
\bibfield{author}{\bibinfo{person}{Allan Kaprow}.}
  \bibinfo{year}{2014}\natexlab{}.
\newblock \showarticletitle{{Happenings in the New York scene}}.
\newblock In \bibinfo{booktitle}{\emph{The Improvisation Studies Reader}}.
  \bibinfo{publisher}{Routledge}, \bibinfo{pages}{254--260}.
\newblock


\bibitem[\protect\citeauthoryear{Kaufman and Beghetto}{Kaufman and
  Beghetto}{2009}]%
        {kaufman2009beyond}
\bibfield{author}{\bibinfo{person}{James~C Kaufman} {and}
  \bibinfo{person}{Ronald~A Beghetto}.} \bibinfo{year}{2009}\natexlab{}.
\newblock \showarticletitle{{Beyond big and little: The four c model of
  creativity.}}
\newblock \bibinfo{journal}{\emph{Review of general psychology}}
  \bibinfo{volume}{13}, \bibinfo{number}{1} (\bibinfo{year}{2009}),
  \bibinfo{pages}{1}.
\newblock


\bibitem[\protect\citeauthoryear{Kerne}{Kerne}{2001}]%
        {kerne2001collagemachine}
\bibfield{author}{\bibinfo{person}{Andruid Kerne}.}
  \bibinfo{year}{2001}\natexlab{}.
\newblock \emph{\bibinfo{title}{CollageMachine: A model of interface ecology}}.
\newblock \bibinfo{thesistype}{Ph.D. Dissertation}. \bibinfo{school}{New York
  University, Graduate School of Arts and Science}.
\newblock


\bibitem[\protect\citeauthoryear{Kerne}{Kerne}{2002}]%
        {kerne2002interface}
\bibfield{author}{\bibinfo{person}{Andruid Kerne}.}
  \bibinfo{year}{2002}\natexlab{}.
\newblock \showarticletitle{Interface ecosystem, the fundamental unit of
  information age ecology}. In \bibinfo{booktitle}{\emph{Proceedings of the
  29th International Conference on Computer Graphics and Interactive
  Techniques. Electronic Art and Animation Catalog.}}
  \bibinfo{pages}{142--145}.
\newblock


\bibitem[\protect\citeauthoryear{Kerne, Lupfer, Linder, Qu, Valdez, Jain,
  Keith, Carrasco, Vanegas, and Billingsley}{Kerne et~al\mbox{.}}{2017}]%
        {kerne2017strategies}
\bibfield{author}{\bibinfo{person}{Andruid Kerne}, \bibinfo{person}{Nic
  Lupfer}, \bibinfo{person}{Rhema Linder}, \bibinfo{person}{Yin Qu},
  \bibinfo{person}{Alyssa Valdez}, \bibinfo{person}{Ajit Jain},
  \bibinfo{person}{Kade Keith}, \bibinfo{person}{Matthew Carrasco},
  \bibinfo{person}{Jorge Vanegas}, {and} \bibinfo{person}{Andrew Billingsley}.}
  \bibinfo{year}{2017}\natexlab{}.
\newblock \showarticletitle{Strategies of Free-Form Web Curation: Processes of
  Creative Engagement with Prior Work}. In
  \bibinfo{booktitle}{\emph{Proceedings of the 2017 ACM SIGCHI Conference on
  Creativity and Cognition}}. \bibinfo{pages}{380--392}.
\newblock


\bibitem[\protect\citeauthoryear{Kerne, Webb, Smith, Linder, Lupfer, Qu,
  Moeller, and Damaraju}{Kerne et~al\mbox{.}}{2014}]%
        {kerne2014using}
\bibfield{author}{\bibinfo{person}{Andruid Kerne}, \bibinfo{person}{Andrew~M
  Webb}, \bibinfo{person}{Steven~M Smith}, \bibinfo{person}{Rhema Linder},
  \bibinfo{person}{Nic Lupfer}, \bibinfo{person}{Yin Qu}, \bibinfo{person}{Jon
  Moeller}, {and} \bibinfo{person}{Sashikanth Damaraju}.}
  \bibinfo{year}{2014}\natexlab{}.
\newblock \showarticletitle{Using metrics of curation to evaluate
  information-based ideation}.
\newblock \bibinfo{journal}{\emph{ACM Transactions on Computer-Human
  Interaction (TOCHI)}} \bibinfo{volume}{21}, \bibinfo{number}{3}
  (\bibinfo{year}{2014}), \bibinfo{pages}{1--48}.
\newblock


\bibitem[\protect\citeauthoryear{Kim, Agrawala, and Bernstein}{Kim
  et~al\mbox{.}}{2017}]%
        {kim2017mosaic}
\bibfield{author}{\bibinfo{person}{Joy Kim}, \bibinfo{person}{Maneesh
  Agrawala}, {and} \bibinfo{person}{Michael~S Bernstein}.}
  \bibinfo{year}{2017}\natexlab{}.
\newblock \showarticletitle{{Mosaic: designing online creative communities for
  sharing works-in-progress}}. In \bibinfo{booktitle}{\emph{Proceedings of the
  2017 ACM Conference on Computer Supported Cooperative Work and Social
  Computing}}. ACM, \bibinfo{pages}{246--258}.
\newblock


\bibitem[\protect\citeauthoryear{Kittur, Yu, Hope, Chan, Lifshitz-Assaf, Gilon,
  Ng, Kraut, and Shahaf}{Kittur et~al\mbox{.}}{2019}]%
        {kittur2019scaling}
\bibfield{author}{\bibinfo{person}{Aniket Kittur}, \bibinfo{person}{Lixiu Yu},
  \bibinfo{person}{Tom Hope}, \bibinfo{person}{Joel Chan},
  \bibinfo{person}{Hila Lifshitz-Assaf}, \bibinfo{person}{Karni Gilon},
  \bibinfo{person}{Felicia Ng}, \bibinfo{person}{Robert~E Kraut}, {and}
  \bibinfo{person}{Dafna Shahaf}.} \bibinfo{year}{2019}\natexlab{}.
\newblock \showarticletitle{Scaling up analogical innovation with crowds and
  AI}.
\newblock \bibinfo{journal}{\emph{Proceedings of the National Academy of
  Sciences}} \bibinfo{volume}{116}, \bibinfo{number}{6} (\bibinfo{year}{2019}),
  \bibinfo{pages}{1870--1877}.
\newblock


\bibitem[\protect\citeauthoryear{Kohn and Smith}{Kohn and Smith}{2011}]%
        {kohn2011collaborative}
\bibfield{author}{\bibinfo{person}{Nicholas~W Kohn} {and}
  \bibinfo{person}{Steven~M Smith}.} \bibinfo{year}{2011}\natexlab{}.
\newblock \showarticletitle{{Collaborative fixation: Effects of others' ideas
  on brainstorming}}.
\newblock \bibinfo{journal}{\emph{Applied Cognitive Psychology}}
  \bibinfo{volume}{25}, \bibinfo{number}{3} (\bibinfo{year}{2011}),
  \bibinfo{pages}{359--371}.
\newblock


\bibitem[\protect\citeauthoryear{Kolko}{Kolko}{2012}]%
        {kolko2012transformative}
\bibfield{author}{\bibinfo{person}{Jon Kolko}.}
  \bibinfo{year}{2012}\natexlab{}.
\newblock \showarticletitle{{Transformative learning in the design studio.}}
\newblock \bibinfo{journal}{\emph{interactions}} \bibinfo{volume}{19},
  \bibinfo{number}{6} (\bibinfo{year}{2012}), \bibinfo{pages}{82--83}.
\newblock


\bibitem[\protect\citeauthoryear{Krause, Garncarz, Song, Gerber, Bailey, and
  Dow}{Krause et~al\mbox{.}}{2017}]%
        {krause2017critique}
\bibfield{author}{\bibinfo{person}{Markus Krause}, \bibinfo{person}{Tom
  Garncarz}, \bibinfo{person}{JiaoJiao Song}, \bibinfo{person}{Elizabeth~M
  Gerber}, \bibinfo{person}{Brian~P Bailey}, {and} \bibinfo{person}{Steven~P
  Dow}.} \bibinfo{year}{2017}\natexlab{}.
\newblock \showarticletitle{Critique style guide: Improving crowdsourced design
  feedback with a natural language model}. In
  \bibinfo{booktitle}{\emph{Proceedings of the 2017 CHI Conference on Human
  Factors in Computing Systems}}. \bibinfo{pages}{4627--4639}.
\newblock


\bibitem[\protect\citeauthoryear{Kreijns, Kirschner, and Jochems}{Kreijns
  et~al\mbox{.}}{2003}]%
        {kreijns2003identifying}
\bibfield{author}{\bibinfo{person}{Karel Kreijns}, \bibinfo{person}{Paul~A
  Kirschner}, {and} \bibinfo{person}{Wim Jochems}.}
  \bibinfo{year}{2003}\natexlab{}.
\newblock \showarticletitle{{Identifying the pitfalls for social interaction in
  computer-supported collaborative learning environments: a review of the
  research}}.
\newblock \bibinfo{journal}{\emph{Computers in human behavior}}
  \bibinfo{volume}{19}, \bibinfo{number}{3} (\bibinfo{year}{2003}),
  \bibinfo{pages}{335--353}.
\newblock


\bibitem[\protect\citeauthoryear{Kyng and Greenbaum}{Kyng and
  Greenbaum}{1991}]%
        {kyng1991design}
\bibfield{author}{\bibinfo{person}{Morten Kyng} {and} \bibinfo{person}{Joan
  Greenbaum}.} \bibinfo{year}{1991}\natexlab{}.
\newblock \bibinfo{booktitle}{\emph{{Design at work: Cooperative design of
  computer systems}}}.
\newblock \bibinfo{publisher}{Erlbaum, Lawrence, Associates}.
\newblock


\bibitem[\protect\citeauthoryear{Lande and Leifer}{Lande and Leifer}{2010}]%
        {lande2010difficulties}
\bibfield{author}{\bibinfo{person}{Micah Lande} {and} \bibinfo{person}{Larry
  Leifer}.} \bibinfo{year}{2010}\natexlab{}.
\newblock \showarticletitle{{Difficulties student engineers face designing the
  future}}.
\newblock \bibinfo{journal}{\emph{International Journal of Engineering
  Education}} \bibinfo{volume}{26}, \bibinfo{number}{2} (\bibinfo{year}{2010}),
  \bibinfo{pages}{271}.
\newblock


\bibitem[\protect\citeauthoryear{Lave and Wenger}{Lave and Wenger}{1991}]%
        {lave1991situated}
\bibfield{author}{\bibinfo{person}{Jean Lave} {and} \bibinfo{person}{Etienne
  Wenger}.} \bibinfo{year}{1991}\natexlab{}.
\newblock \bibinfo{booktitle}{\emph{{Situated learning: Legitimate peripheral
  participation}}}.
\newblock \bibinfo{publisher}{Cambridge university press}.
\newblock


\bibitem[\protect\citeauthoryear{Lim}{Lim}{2018}]%
        {lim2018design}
\bibfield{author}{\bibinfo{person}{Hajin Lim}.}
  \bibinfo{year}{2018}\natexlab{}.
\newblock \showarticletitle{Design for Computer-Mediated Multilingual
  Communication with AI Support}. In \bibinfo{booktitle}{\emph{Companion of the
  2018 ACM Conference on Computer Supported Cooperative Work and Social
  Computing}}. \bibinfo{pages}{93--96}.
\newblock


\bibitem[\protect\citeauthoryear{Linsey, Wood, and Markman}{Linsey
  et~al\mbox{.}}{2009}]%
        {linsey2009increasing}
\bibfield{author}{\bibinfo{person}{Julie~S Linsey}, \bibinfo{person}{Kristin~L
  Wood}, {and} \bibinfo{person}{Arthur~B Markman}.}
  \bibinfo{year}{2009}\natexlab{}.
\newblock \showarticletitle{{Increasing innovation: presentation and evaluation
  of the wordtree design-by-analogy method}}. In \bibinfo{booktitle}{\emph{ASME
  2008 International Design Engineering Technical Conferences and Computers and
  Information in Engineering Conference}}. American Society of Mechanical
  Engineers Digital Collection, \bibinfo{pages}{21--32}.
\newblock


\bibitem[\protect\citeauthoryear{Lippard}{Lippard}{1997}]%
        {lippard1997six}
\bibfield{author}{\bibinfo{person}{Lucy~R Lippard}.}
  \bibinfo{year}{1997}\natexlab{}.
\newblock \bibinfo{booktitle}{\emph{{Six years: the dematerialization of the
  art object from 1966 to 1972}}}. Vol.~\bibinfo{volume}{364}.
\newblock \bibinfo{publisher}{Univ of California Press}.
\newblock


\bibitem[\protect\citeauthoryear{Loughry, Ohland, and Woehr}{Loughry
  et~al\mbox{.}}{2014}]%
        {loughry2014assessing}
\bibfield{author}{\bibinfo{person}{Misty~L Loughry}, \bibinfo{person}{Matthew~W
  Ohland}, {and} \bibinfo{person}{David~J Woehr}.}
  \bibinfo{year}{2014}\natexlab{}.
\newblock \showarticletitle{Assessing teamwork skills for assurance of learning
  using CATME team tools}.
\newblock \bibinfo{journal}{\emph{Journal of Marketing Education}}
  \bibinfo{volume}{36}, \bibinfo{number}{1} (\bibinfo{year}{2014}),
  \bibinfo{pages}{5--19}.
\newblock


\bibitem[\protect\citeauthoryear{Luck}{Luck}{2003}]%
        {luck2003dialogue}
\bibfield{author}{\bibinfo{person}{Rachael Luck}.}
  \bibinfo{year}{2003}\natexlab{}.
\newblock \showarticletitle{Dialogue in participatory design}.
\newblock \bibinfo{journal}{\emph{Design studies}} \bibinfo{volume}{24},
  \bibinfo{number}{6} (\bibinfo{year}{2003}), \bibinfo{pages}{523--535}.
\newblock


\bibitem[\protect\citeauthoryear{Lupfer, Fowler, Valdez, Webb, Merrill, Newman,
  and Kerne}{Lupfer et~al\mbox{.}}{2018}]%
        {lupfer2018multiscale}
\bibfield{author}{\bibinfo{person}{Nic Lupfer}, \bibinfo{person}{Hannah
  Fowler}, \bibinfo{person}{Alyssa Valdez}, \bibinfo{person}{Andrew Webb},
  \bibinfo{person}{Jeremy Merrill}, \bibinfo{person}{Galen Newman}, {and}
  \bibinfo{person}{Andruid Kerne}.} \bibinfo{year}{2018}\natexlab{}.
\newblock \showarticletitle{{Multiscale Design Strategies in a Landscape
  Architecture Classroom}}. In \bibinfo{booktitle}{\emph{Proceedings of the
  2018 on Designing Interactive Systems Conference 2018}}. ACM,
  \bibinfo{pages}{1081--1093}.
\newblock


\bibitem[\protect\citeauthoryear{Lupfer, Kerne, Linder, Fowler, Rajanna,
  Carrasco, and Valdez}{Lupfer et~al\mbox{.}}{2019}]%
        {lupfer2019multiscale}
\bibfield{author}{\bibinfo{person}{Nic Lupfer}, \bibinfo{person}{Andruid
  Kerne}, \bibinfo{person}{Rhema Linder}, \bibinfo{person}{Hannah Fowler},
  \bibinfo{person}{Vijay Rajanna}, \bibinfo{person}{Matthew Carrasco}, {and}
  \bibinfo{person}{Alyssa Valdez}.} \bibinfo{year}{2019}\natexlab{}.
\newblock \showarticletitle{{Multiscale Design Curation: Supporting Computer
  Science Students' Iterative and Reflective Creative Processes}}. In
  \bibinfo{booktitle}{\emph{Proceedings of the 2019 on Creativity and
  Cognition}}. ACM, \bibinfo{pages}{233--245}.
\newblock


\bibitem[\protect\citeauthoryear{Lupfer, Kerne, Webb, and Linder}{Lupfer
  et~al\mbox{.}}{2016}]%
        {lupfer2016patterns}
\bibfield{author}{\bibinfo{person}{Nic Lupfer}, \bibinfo{person}{Andruid
  Kerne}, \bibinfo{person}{Andrew~M Webb}, {and} \bibinfo{person}{Rhema
  Linder}.} \bibinfo{year}{2016}\natexlab{}.
\newblock \showarticletitle{Patterns of free-form curation: Visual thinking
  with web content}. In \bibinfo{booktitle}{\emph{Proceedings of the 24th ACM
  international conference on Multimedia}}. \bibinfo{pages}{12--21}.
\newblock


\bibitem[\protect\citeauthoryear{Luther, Tolentino, Wu, Pavel, Bailey,
  Agrawala, Hartmann, and Dow}{Luther et~al\mbox{.}}{2015}]%
        {luther2015structuring}
\bibfield{author}{\bibinfo{person}{Kurt Luther}, \bibinfo{person}{Jari-Lee
  Tolentino}, \bibinfo{person}{Wei Wu}, \bibinfo{person}{Amy Pavel},
  \bibinfo{person}{Brian~P Bailey}, \bibinfo{person}{Maneesh Agrawala},
  \bibinfo{person}{Bj{\"{o}}rn Hartmann}, {and} \bibinfo{person}{Steven~P
  Dow}.} \bibinfo{year}{2015}\natexlab{}.
\newblock \showarticletitle{{Structuring, aggregating, and evaluating
  crowdsourced design critique}}. In \bibinfo{booktitle}{\emph{Proceedings of
  the 18th ACM Conference on Computer Supported Cooperative Work {\&} Social
  Computing}}. ACM, \bibinfo{pages}{473--485}.
\newblock


\bibitem[\protect\citeauthoryear{Mackeprang, Khiat, and
  M{\"{u}}ller-Birn}{Mackeprang et~al\mbox{.}}{2018}]%
        {mackeprang2018concept}
\bibfield{author}{\bibinfo{person}{Maximilian Mackeprang},
  \bibinfo{person}{Abderrahmane Khiat}, {and} \bibinfo{person}{Claudia
  M{\"{u}}ller-Birn}.} \bibinfo{year}{2018}\natexlab{}.
\newblock \showarticletitle{{Concept Validation during Collaborative Ideation
  and Its Effect on Ideation Outcome}}. In \bibinfo{booktitle}{\emph{CHI
  Extended Abstracts}}. ACM, \bibinfo{pages}{1--6}.
\newblock


\bibitem[\protect\citeauthoryear{Marshall and Shipman~III}{Marshall and
  Shipman~III}{1995}]%
        {marshall1995spatial}
\bibfield{author}{\bibinfo{person}{Catherine~C Marshall} {and}
  \bibinfo{person}{Frank~M Shipman~III}.} \bibinfo{year}{1995}\natexlab{}.
\newblock \showarticletitle{Spatial hypertext: designing for change}.
\newblock \bibinfo{journal}{\emph{Commun. ACM}} \bibinfo{volume}{38},
  \bibinfo{number}{8} (\bibinfo{year}{1995}), \bibinfo{pages}{88--97}.
\newblock


\bibitem[\protect\citeauthoryear{Mayer and Moreno}{Mayer and Moreno}{2003}]%
        {mayer2003nine}
\bibfield{author}{\bibinfo{person}{Richard~E Mayer} {and}
  \bibinfo{person}{Roxana Moreno}.} \bibinfo{year}{2003}\natexlab{}.
\newblock \showarticletitle{Nine ways to reduce cognitive load in multimedia
  learning}.
\newblock \bibinfo{journal}{\emph{Educational psychologist}}
  \bibinfo{volume}{38}, \bibinfo{number}{1} (\bibinfo{year}{2003}),
  \bibinfo{pages}{43--52}.
\newblock


\bibitem[\protect\citeauthoryear{McDonald, Rich, and Gubler}{McDonald
  et~al\mbox{.}}{2019}]%
        {mcdonald2019perceived}
\bibfield{author}{\bibinfo{person}{Jason~K McDonald}, \bibinfo{person}{Peter~J
  Rich}, {and} \bibinfo{person}{Nicholas~B Gubler}.}
  \bibinfo{year}{2019}\natexlab{}.
\newblock \showarticletitle{The perceived value of informal, peer critique in
  the instructional design studio}.
\newblock \bibinfo{journal}{\emph{TechTrends}} \bibinfo{volume}{63},
  \bibinfo{number}{2} (\bibinfo{year}{2019}), \bibinfo{pages}{149--159}.
\newblock


\bibitem[\protect\citeauthoryear{McDonnell}{McDonnell}{2016}]%
        {mcdonnell2016scaffolding}
\bibfield{author}{\bibinfo{person}{Janet McDonnell}.}
  \bibinfo{year}{2016}\natexlab{}.
\newblock \showarticletitle{{Scaffolding practices: A study of design
  practitioner engagement in design education}}.
\newblock \bibinfo{journal}{\emph{Design studies}}  \bibinfo{volume}{45}
  (\bibinfo{year}{2016}), \bibinfo{pages}{9--29}.
\newblock


\bibitem[\protect\citeauthoryear{Mednick}{Mednick}{1962}]%
        {mednick1962associative}
\bibfield{author}{\bibinfo{person}{Sarnoff Mednick}.}
  \bibinfo{year}{1962}\natexlab{}.
\newblock \showarticletitle{{The associative basis of the creative process.}}
\newblock \bibinfo{journal}{\emph{Psychological review}} \bibinfo{volume}{69},
  \bibinfo{number}{3} (\bibinfo{year}{1962}), \bibinfo{pages}{220}.
\newblock


\bibitem[\protect\citeauthoryear{Meggs}{Meggs}{1992}]%
        {meggs1992type}
\bibfield{author}{\bibinfo{person}{Philip~B Meggs}.}
  \bibinfo{year}{1992}\natexlab{}.
\newblock \bibinfo{booktitle}{\emph{{Type and image: The language of graphic
  design}}}.
\newblock \bibinfo{publisher}{John Wiley {\&} Sons}.
\newblock


\bibitem[\protect\citeauthoryear{Michel, Lavou{\'{e}}, George, and Ji}{Michel
  et~al\mbox{.}}{2017}]%
        {michel2017supporting}
\bibfield{author}{\bibinfo{person}{Christine Michel}, \bibinfo{person}{Elise
  Lavou{\'{e}}}, \bibinfo{person}{S{\'{e}}bastien George}, {and}
  \bibinfo{person}{Min Ji}.} \bibinfo{year}{2017}\natexlab{}.
\newblock \showarticletitle{{Supporting awareness and self-regulation in
  project-based learning through personalized dashboards}}.
\newblock \bibinfo{journal}{\emph{International Journal of Technology Enhanced
  Learning}} \bibinfo{volume}{9}, \bibinfo{number}{2--3}
  (\bibinfo{year}{2017}).
\newblock


\bibitem[\protect\citeauthoryear{Nelson}{Nelson}{1990}]%
        {nelson1990xanadu}
\bibfield{author}{\bibinfo{person}{Ted Nelson}.}
  \bibinfo{year}{1990}\natexlab{}.
\newblock \showarticletitle{On the Xanadu project}.
\newblock \bibinfo{journal}{\emph{BYTE Magazine}} \bibinfo{volume}{15},
  \bibinfo{number}{9} (\bibinfo{year}{1990}), \bibinfo{pages}{298--299}.
\newblock


\bibitem[\protect\citeauthoryear{Nelson}{Nelson}{1993}]%
        {nelson1993literary}
\bibfield{author}{\bibinfo{person}{Ted Nelson}.}
  \bibinfo{year}{1993}\natexlab{}.
\newblock \bibinfo{booktitle}{\emph{Literary machines 93.1}}.
\newblock \bibinfo{publisher}{Mindful Pr}.
\newblock


\bibitem[\protect\citeauthoryear{Ngoon, Fraser, Weingarten, Dontcheva, and
  Klemmer}{Ngoon et~al\mbox{.}}{2018}]%
        {ngoon2018interactive}
\bibfield{author}{\bibinfo{person}{Tricia~J Ngoon}, \bibinfo{person}{C~Ailie
  Fraser}, \bibinfo{person}{Ariel~S Weingarten}, \bibinfo{person}{Mira
  Dontcheva}, {and} \bibinfo{person}{Scott Klemmer}.}
  \bibinfo{year}{2018}\natexlab{}.
\newblock \showarticletitle{Interactive guidance techniques for improving
  creative feedback}. In \bibinfo{booktitle}{\emph{Proceedings of the 2018 CHI
  Conference on Human Factors in Computing Systems}}. \bibinfo{pages}{1--11}.
\newblock


\bibitem[\protect\citeauthoryear{of~Sciences, Engineering, and
  Medicine}{of~Sciences et~al\mbox{.}}{2018}]%
        {national2018workforce}
\bibfield{author}{\bibinfo{person}{National~Academies of Sciences},
  \bibinfo{person}{Engineering}, {and} \bibinfo{person}{Medicine}.}
  \bibinfo{year}{2018}\natexlab{}.
\newblock \bibinfo{booktitle}{\emph{Workforce development and intelligence
  analysis for national security purposes: Proceedings of a workshop}}.
\newblock \bibinfo{publisher}{National Academies Press}.
\newblock


\bibitem[\protect\citeauthoryear{Oh, Ishizaki, Gross, and Do}{Oh
  et~al\mbox{.}}{2013}]%
        {oh2013theoretical}
\bibfield{author}{\bibinfo{person}{Yeonjoo Oh}, \bibinfo{person}{Suguru
  Ishizaki}, \bibinfo{person}{Mark~D Gross}, {and} \bibinfo{person}{Ellen
  Yi-Luen Do}.} \bibinfo{year}{2013}\natexlab{}.
\newblock \showarticletitle{{A theoretical framework of design critiquing in
  architecture studios}}.
\newblock \bibinfo{journal}{\emph{Design Studies}} \bibinfo{volume}{34},
  \bibinfo{number}{3} (\bibinfo{year}{2013}), \bibinfo{pages}{302--325}.
\newblock


\bibitem[\protect\citeauthoryear{Ordonez, Kulkarni, and Berg}{Ordonez
  et~al\mbox{.}}{2011}]%
        {ordonez2011im2text}
\bibfield{author}{\bibinfo{person}{Vicente Ordonez}, \bibinfo{person}{Girish
  Kulkarni}, {and} \bibinfo{person}{Tamara~L Berg}.}
  \bibinfo{year}{2011}\natexlab{}.
\newblock \showarticletitle{Im2text: Describing images using 1 million
  captioned photographs}. In \bibinfo{booktitle}{\emph{Advances in neural
  information processing systems}}. \bibinfo{pages}{1143--1151}.
\newblock


\bibitem[\protect\citeauthoryear{Osberger and Maeder}{Osberger and
  Maeder}{1998}]%
        {osberger1998automatic}
\bibfield{author}{\bibinfo{person}{Wilfried Osberger} {and}
  \bibinfo{person}{Anthony~J Maeder}.} \bibinfo{year}{1998}\natexlab{}.
\newblock \showarticletitle{{Automatic identification of perceptually important
  regions in an image}}. In \bibinfo{booktitle}{\emph{Proceedings. Fourteenth
  International Conference on Pattern Recognition (Cat. No. 98EX170)}},
  Vol.~\bibinfo{volume}{1}. IEEE, \bibinfo{pages}{701--704}.
\newblock


\bibitem[\protect\citeauthoryear{Osmond and Tovey}{Osmond and Tovey}{2015}]%
        {osmond2015threshold}
\bibfield{author}{\bibinfo{person}{Jane Osmond} {and} \bibinfo{person}{Michael
  Tovey}.} \bibinfo{year}{2015}\natexlab{}.
\newblock \showarticletitle{{The Threshold of Uncertainty in Teaching Design.}}
\newblock \bibinfo{journal}{\emph{Design and Technology Education}}
  \bibinfo{volume}{20}, \bibinfo{number}{2} (\bibinfo{year}{2015}),
  \bibinfo{pages}{50--57}.
\newblock


\bibitem[\protect\citeauthoryear{Oulasvirta, {De Pascale}, Koch, Langerak,
  Jokinen, Todi, Laine, Kristhombuge, Zhu, Miniukovich, and Others}{Oulasvirta
  et~al\mbox{.}}{2018}]%
        {oulasvirta2018aalto}
\bibfield{author}{\bibinfo{person}{Antti Oulasvirta}, \bibinfo{person}{Samuli
  {De Pascale}}, \bibinfo{person}{Janin Koch}, \bibinfo{person}{Thomas
  Langerak}, \bibinfo{person}{Jussi Jokinen}, \bibinfo{person}{Kashyap Todi},
  \bibinfo{person}{Markku Laine}, \bibinfo{person}{Manoj Kristhombuge},
  \bibinfo{person}{Yuxi Zhu}, \bibinfo{person}{Aliaksei Miniukovich}, {and}
  \bibinfo{person}{Others}.} \bibinfo{year}{2018}\natexlab{}.
\newblock \showarticletitle{{Aalto Interface Metrics (AIM) A Service and
  Codebase for Computational GUI Evaluation}}. In \bibinfo{booktitle}{\emph{The
  31st Annual ACM Symposium on User Interface Software and Technology Adjunct
  Proceedings}}. \bibinfo{pages}{16--19}.
\newblock


\bibitem[\protect\citeauthoryear{Pan and Yang}{Pan and Yang}{2009}]%
        {pan2009survey}
\bibfield{author}{\bibinfo{person}{Sinno~Jialin Pan} {and}
  \bibinfo{person}{Qiang Yang}.} \bibinfo{year}{2009}\natexlab{}.
\newblock \showarticletitle{{A survey on transfer learning}}.
\newblock \bibinfo{journal}{\emph{IEEE Transactions on knowledge and data
  engineering}} \bibinfo{volume}{22}, \bibinfo{number}{10}
  (\bibinfo{year}{2009}), \bibinfo{pages}{1345--1359}.
\newblock


\bibitem[\protect\citeauthoryear{Papadopoulos, Sritanyaratana, and
  Klemmer}{Papadopoulos et~al\mbox{.}}{2014}]%
        {papadopoulos2014community}
\bibfield{author}{\bibinfo{person}{Kathryn Papadopoulos},
  \bibinfo{person}{Lalida Sritanyaratana}, {and} \bibinfo{person}{Scott~R
  Klemmer}.} \bibinfo{year}{2014}\natexlab{}.
\newblock \showarticletitle{{Community TAs scale high-touch learning, provide
  student-staff brokering, and build esprit de corps}}. In
  \bibinfo{booktitle}{\emph{Proceedings of the first ACM conference on
  Learning@ scale conference}}. ACM, \bibinfo{pages}{163--164}.
\newblock


\bibitem[\protect\citeauthoryear{Pennington, Socher, and Manning}{Pennington
  et~al\mbox{.}}{2014}]%
        {pennington2014glove}
\bibfield{author}{\bibinfo{person}{Jeffrey Pennington},
  \bibinfo{person}{Richard Socher}, {and} \bibinfo{person}{Christopher
  Manning}.} \bibinfo{year}{2014}\natexlab{}.
\newblock \showarticletitle{{Glove: Global vectors for word representation}}.
  In \bibinfo{booktitle}{\emph{Proceedings of the 2014 conference on empirical
  methods in natural language processing (EMNLP)}}.
  \bibinfo{pages}{1532--1543}.
\newblock


\bibitem[\protect\citeauthoryear{Reinecke, Yeh, Miratrix, Mardiko, Zhao, Liu,
  and Gajos}{Reinecke et~al\mbox{.}}{2013}]%
        {reinecke2013predicting}
\bibfield{author}{\bibinfo{person}{Katharina Reinecke}, \bibinfo{person}{Tom
  Yeh}, \bibinfo{person}{Luke Miratrix}, \bibinfo{person}{Rahmatri Mardiko},
  \bibinfo{person}{Yuechen Zhao}, \bibinfo{person}{Jenny Liu}, {and}
  \bibinfo{person}{Krzysztof~Z Gajos}.} \bibinfo{year}{2013}\natexlab{}.
\newblock \showarticletitle{{Predicting users' first impressions of website
  aesthetics with a quantification of perceived visual complexity and
  colorfulness}}. In \bibinfo{booktitle}{\emph{Proc. CHI}}. ACM,
  \bibinfo{pages}{2049--2058}.
\newblock


\bibitem[\protect\citeauthoryear{Rikakis, Kelliher, Huang, and
  Sundaram}{Rikakis et~al\mbox{.}}{2018}]%
        {rikakis2018progressive}
\bibfield{author}{\bibinfo{person}{Thanassis Rikakis}, \bibinfo{person}{Aisling
  Kelliher}, \bibinfo{person}{Jia-Bin Huang}, {and} \bibinfo{person}{Hari
  Sundaram}.} \bibinfo{year}{2018}\natexlab{}.
\newblock \showarticletitle{{Progressive cyber-human intelligence for social
  good}}.
\newblock \bibinfo{journal}{\emph{Interactions}} \bibinfo{volume}{25},
  \bibinfo{number}{4} (\bibinfo{year}{2018}), \bibinfo{pages}{52--56}.
\newblock


\bibitem[\protect\citeauthoryear{Rittel and Webber}{Rittel and Webber}{1973}]%
        {rittel1973dilemmas}
\bibfield{author}{\bibinfo{person}{Horst W~J Rittel} {and}
  \bibinfo{person}{Melvin~M Webber}.} \bibinfo{year}{1973}\natexlab{}.
\newblock \showarticletitle{{Dilemmas in a general theory of planning}}.
\newblock \bibinfo{journal}{\emph{Policy sciences}} \bibinfo{volume}{4},
  \bibinfo{number}{2} (\bibinfo{year}{1973}), \bibinfo{pages}{155--169}.
\newblock


\bibitem[\protect\citeauthoryear{Sanders and Stappers}{Sanders and
  Stappers}{2008}]%
        {sanders2008co}
\bibfield{author}{\bibinfo{person}{Elizabeth B-N Sanders} {and}
  \bibinfo{person}{Pieter~Jan Stappers}.} \bibinfo{year}{2008}\natexlab{}.
\newblock \showarticletitle{Co-creation and the new landscapes of design}.
\newblock \bibinfo{journal}{\emph{Co-design}} \bibinfo{volume}{4},
  \bibinfo{number}{1} (\bibinfo{year}{2008}), \bibinfo{pages}{5--18}.
\newblock


\bibitem[\protect\citeauthoryear{Sanders and Stappers}{Sanders and
  Stappers}{2012}]%
        {sanders2012convivial}
\bibfield{author}{\bibinfo{person}{Elizabeth B-N Sanders} {and}
  \bibinfo{person}{Pieter~Jan Stappers}.} \bibinfo{year}{2012}\natexlab{}.
\newblock \bibinfo{booktitle}{\emph{{Convivial toolbox: Generative research for
  the front end of design}}}.
\newblock \bibinfo{publisher}{BIS Amsterdam}.
\newblock


\bibitem[\protect\citeauthoryear{Sankey}{Sankey}{2003}]%
        {sankey2003scientific}
\bibfield{author}{\bibinfo{person}{Howard Sankey}.}
  \bibinfo{year}{2003}\natexlab{}.
\newblock \showarticletitle{{Scientific realism and the God's eye point of
  view}}.
\newblock \bibinfo{journal}{\emph{Epistemologia}} \bibinfo{volume}{27},
  \bibinfo{number}{2} (\bibinfo{year}{2003}), \bibinfo{pages}{211--26}.
\newblock


\bibitem[\protect\citeauthoryear{Sawyer}{Sawyer}{2017}]%
        {sawyer2017teaching}
\bibfield{author}{\bibinfo{person}{R~Keith Sawyer}.}
  \bibinfo{year}{2017}\natexlab{}.
\newblock \showarticletitle{{Teaching creativity in art and design studio
  classes: A systematic literature review}}.
\newblock \bibinfo{journal}{\emph{Educational research review}}
  \bibinfo{volume}{22} (\bibinfo{year}{2017}), \bibinfo{pages}{99--113}.
\newblock


\bibitem[\protect\citeauthoryear{Sawyer}{Sawyer}{2018}]%
        {sawyer2018teaching}
\bibfield{author}{\bibinfo{person}{R~Keith Sawyer}.}
  \bibinfo{year}{2018}\natexlab{}.
\newblock \showarticletitle{{Teaching and learning how to create in schools of
  art and design}}.
\newblock \bibinfo{journal}{\emph{Journal of the Learning Sciences}}
  \bibinfo{volume}{27}, \bibinfo{number}{1} (\bibinfo{year}{2018}),
  \bibinfo{pages}{137--181}.
\newblock


\bibitem[\protect\citeauthoryear{Schuler and Namioka}{Schuler and
  Namioka}{1993}]%
        {schuler1993participatory}
\bibfield{author}{\bibinfo{person}{Douglas Schuler} {and} \bibinfo{person}{Aki
  Namioka}.} \bibinfo{year}{1993}\natexlab{}.
\newblock \bibinfo{booktitle}{\emph{{Participatory design: Principles and
  practices}}}.
\newblock \bibinfo{publisher}{CRC Press}.
\newblock


\bibitem[\protect\citeauthoryear{Seitz, {of modern art (New York}, {of
  Contemporary Arts (Tex.).}, and {of Modern Art (Calif.).}}{Seitz
  et~al\mbox{.}}{1961}]%
        {seitz1961art}
\bibfield{author}{\bibinfo{person}{William~Chapin Seitz}, \bibinfo{person}{NY).
  {of modern art (New York}}, \bibinfo{person}{Dallas~Museum {of Contemporary
  Arts (Tex.).}}, {and} \bibinfo{person}{San Francisco~Museum {of Modern Art
  (Calif.).}}} \bibinfo{year}{1961}\natexlab{}.
\newblock \bibinfo{booktitle}{\emph{{The art of assemblage}}}.
\newblock \bibinfo{publisher}{Museum of Modern Art New York}.
\newblock


\bibitem[\protect\citeauthoryear{Shah, Smith, and Vargas-Hernandez}{Shah
  et~al\mbox{.}}{2003}]%
        {shah2003metrics}
\bibfield{author}{\bibinfo{person}{Jami~J Shah}, \bibinfo{person}{Steve~M
  Smith}, {and} \bibinfo{person}{Noe Vargas-Hernandez}.}
  \bibinfo{year}{2003}\natexlab{}.
\newblock \showarticletitle{{Metrics for measuring ideation effectiveness}}.
\newblock \bibinfo{journal}{\emph{Design studies}} \bibinfo{volume}{24},
  \bibinfo{number}{2} (\bibinfo{year}{2003}), \bibinfo{pages}{111--134}.
\newblock


\bibitem[\protect\citeauthoryear{Shneiderman}{Shneiderman}{2020}]%
        {shneiderman2020human}
\bibfield{author}{\bibinfo{person}{Ben Shneiderman}.}
  \bibinfo{year}{2020}\natexlab{}.
\newblock \showarticletitle{Human-Centered Artificial Intelligence: Reliable,
  Safe \& Trustworthy}.
\newblock \bibinfo{journal}{\emph{International Journal of Human--Computer
  Interaction}} (\bibinfo{year}{2020}), \bibinfo{pages}{1--10}.
\newblock


\bibitem[\protect\citeauthoryear{Shreeve, Sims, and Trowler}{Shreeve
  et~al\mbox{.}}{2010}]%
        {shreeve2010kind}
\bibfield{author}{\bibinfo{person}{Alison Shreeve}, \bibinfo{person}{Ellen
  Sims}, {and} \bibinfo{person}{Paul Trowler}.}
  \bibinfo{year}{2010}\natexlab{}.
\newblock \showarticletitle{{`A kind of exchange': learning from art and design
  teaching}}.
\newblock \bibinfo{journal}{\emph{Higher Education Research {\&} Development}}
  \bibinfo{volume}{29}, \bibinfo{number}{2} (\bibinfo{year}{2010}),
  \bibinfo{pages}{125--138}.
\newblock


\bibitem[\protect\citeauthoryear{Siangliulue, Chan, Dow, and Gajos}{Siangliulue
  et~al\mbox{.}}{2016}]%
        {siangliulue2016ideahound}
\bibfield{author}{\bibinfo{person}{Pao Siangliulue}, \bibinfo{person}{Joel
  Chan}, \bibinfo{person}{Steven~P Dow}, {and} \bibinfo{person}{Krzysztof~Z
  Gajos}.} \bibinfo{year}{2016}\natexlab{}.
\newblock \showarticletitle{{IdeaHound: improving large-scale collaborative
  ideation with crowd-powered real-time semantic modeling}}. In
  \bibinfo{booktitle}{\emph{Proc. UIST}}. ACM, \bibinfo{pages}{609--624}.
\newblock


\bibitem[\protect\citeauthoryear{Siemens and Long}{Siemens and Long}{2011}]%
        {siemens2011penetrating}
\bibfield{author}{\bibinfo{person}{George Siemens} {and} \bibinfo{person}{Phil
  Long}.} \bibinfo{year}{2011}\natexlab{}.
\newblock \showarticletitle{{Penetrating the fog: Analytics in learning and
  education.}}
\newblock \bibinfo{journal}{\emph{EDUCAUSE review}} \bibinfo{volume}{46},
  \bibinfo{number}{5} (\bibinfo{year}{2011}), \bibinfo{pages}{30}.
\newblock


\bibitem[\protect\citeauthoryear{Simonton}{Simonton}{1999}]%
        {simonton1999}
\bibfield{author}{\bibinfo{person}{Dean~Keith Simonton}.}
  \bibinfo{year}{1999}\natexlab{}.
\newblock \showarticletitle{Creativity as Blind Variation and Selective
  Retention: Is the Creative Process Darwinian?}
\newblock \bibinfo{journal}{\emph{Psychological Inquiry}} \bibinfo{volume}{10},
  \bibinfo{number}{4} (\bibinfo{year}{1999}).
\newblock
\showISSN{1047840X}
\urldef\tempurl%
\url{http://www.jstor.org/stable/1449455}
\showURL{%
\tempurl}


\bibitem[\protect\citeauthoryear{Sj{\"o}berg and Timpka}{Sj{\"o}berg and
  Timpka}{1998}]%
        {sjoberg1998participatory}
\bibfield{author}{\bibinfo{person}{Cecilia Sj{\"o}berg} {and}
  \bibinfo{person}{Toomas Timpka}.} \bibinfo{year}{1998}\natexlab{}.
\newblock \showarticletitle{Participatory design of information systems in
  health care}.
\newblock \bibinfo{journal}{\emph{Journal of the American Medical Informatics
  Association}} \bibinfo{volume}{5}, \bibinfo{number}{2}
  (\bibinfo{year}{1998}), \bibinfo{pages}{177--183}.
\newblock


\bibitem[\protect\citeauthoryear{Smith, Ward, and Finke}{Smith
  et~al\mbox{.}}{1995}]%
        {smith1995creative}
\bibfield{author}{\bibinfo{person}{Steven~M Smith}, \bibinfo{person}{Thomas~B
  Ward}, {and} \bibinfo{person}{Ronald~A Finke}.}
  \bibinfo{year}{1995}\natexlab{}.
\newblock \bibinfo{booktitle}{\emph{{The creative cognition approach}}}.
\newblock \bibinfo{publisher}{MIT press}.
\newblock


\bibitem[\protect\citeauthoryear{Star and Griesemer}{Star and
  Griesemer}{1989}]%
        {star1989institutional}
\bibfield{author}{\bibinfo{person}{Susan~Leigh Star} {and}
  \bibinfo{person}{James~R Griesemer}.} \bibinfo{year}{1989}\natexlab{}.
\newblock \showarticletitle{{Institutional ecology,translations' and boundary
  objects: Amateurs and professionals in Berkeley's Museum of Vertebrate
  Zoology, 1907-39}}.
\newblock \bibinfo{journal}{\emph{Social studies of science}}
  \bibinfo{volume}{19}, \bibinfo{number}{3} (\bibinfo{year}{1989}),
  \bibinfo{pages}{387--420}.
\newblock


\bibitem[\protect\citeauthoryear{Stolterman}{Stolterman}{2008}]%
        {stolterman2008nature}
\bibfield{author}{\bibinfo{person}{Erik Stolterman}.}
  \bibinfo{year}{2008}\natexlab{}.
\newblock \showarticletitle{{The nature of design practice and implications for
  interaction design research}}.
\newblock \bibinfo{journal}{\emph{International Journal of Design}}
  \bibinfo{volume}{2}, \bibinfo{number}{1} (\bibinfo{year}{2008}).
\newblock


\bibitem[\protect\citeauthoryear{Suchman}{Suchman}{1995}]%
        {suchman1995making}
\bibfield{author}{\bibinfo{person}{Lucy Suchman}.}
  \bibinfo{year}{1995}\natexlab{}.
\newblock \showarticletitle{Making work visible}.
\newblock \bibinfo{journal}{\emph{Commun. ACM}} \bibinfo{volume}{38},
  \bibinfo{number}{9} (\bibinfo{year}{1995}), \bibinfo{pages}{56--64}.
\newblock


\bibitem[\protect\citeauthoryear{Suchman}{Suchman}{1987}]%
        {suchman1987plans}
\bibfield{author}{\bibinfo{person}{Lucy~A Suchman}.}
  \bibinfo{year}{1987}\natexlab{}.
\newblock \bibinfo{booktitle}{\emph{Plans and situated actions: The problem of
  human-machine communication}}.
\newblock \bibinfo{publisher}{Cambridge university press}.
\newblock


\bibitem[\protect\citeauthoryear{Suthers and Verbert}{Suthers and
  Verbert}{2013}]%
        {suthers2013learning}
\bibfield{author}{\bibinfo{person}{Dan Suthers} {and} \bibinfo{person}{Katrien
  Verbert}.} \bibinfo{year}{2013}\natexlab{}.
\newblock \showarticletitle{{Learning analytics as a middle space}}. In
  \bibinfo{booktitle}{\emph{Proceedings of the Third International Conference
  on Learning Analytics and Knowledge}}. ACM, \bibinfo{pages}{1--4}.
\newblock


\bibitem[\protect\citeauthoryear{Sympli}{Sympli}{2020}]%
        {gitForDesigners}
\bibfield{author}{\bibinfo{person}{Sympli}.} \bibinfo{year}{2020}\natexlab{}.
\newblock \bibinfo{title}{{Versions: Git for Designers}}.
\newblock
\newblock
\urldef\tempurl%
\url{https://versions.sympli.io/}
\showURL{%
Retrieved 2020-03-30 from \tempurl}


\bibitem[\protect\citeauthoryear{Tang, Lai, Arthur, and Leung}{Tang
  et~al\mbox{.}}{1999}]%
        {tang1999students}
\bibfield{author}{\bibinfo{person}{Catherine Tang}, \bibinfo{person}{Patrick
  Lai}, \bibinfo{person}{David Arthur}, {and} \bibinfo{person}{Sau~F Leung}.}
  \bibinfo{year}{1999}\natexlab{}.
\newblock \showarticletitle{{How do students prepare for traditional and
  portfolio assessment in a problem-based learning curriculum}}. In
  \bibinfo{booktitle}{\emph{Themes and Variations in PBL: refereed proceedings
  of the 1999 Bi-ennial PBL Conference}}, Vol.~\bibinfo{volume}{1}. Australia
  ProblemBased LearningNetwork (PROBLARC), \bibinfo{pages}{206--217}.
\newblock


\bibitem[\protect\citeauthoryear{Taylor and McCormack}{Taylor and
  McCormack}{2004}]%
        {taylor2004juggling}
\bibfield{author}{\bibinfo{person}{Mary-Jane Taylor} {and}
  \bibinfo{person}{Coralie McCormack}.} \bibinfo{year}{2004}\natexlab{}.
\newblock \showarticletitle{{Juggling cats: investigating effective verbal
  feedback in graphic design critiques}}. In \bibinfo{booktitle}{\emph{A paper
  to The Australian Council of University Art and Design Schools Annual
  Conference}}. \bibinfo{pages}{22--25}.
\newblock


\bibitem[\protect\citeauthoryear{Tufte, Goeler, and Benson}{Tufte
  et~al\mbox{.}}{1990}]%
        {tufte1990envisioning}
\bibfield{author}{\bibinfo{person}{Edward~R Tufte},
  \bibinfo{person}{Nora~Hillman Goeler}, {and} \bibinfo{person}{Richard
  Benson}.} \bibinfo{year}{1990}\natexlab{}.
\newblock \bibinfo{booktitle}{\emph{{Envisioning information}}}.
  Vol.~\bibinfo{volume}{126}.
\newblock \bibinfo{publisher}{Graphics press Cheshire, CT}.
\newblock


\bibitem[\protect\citeauthoryear{Turnbull and Watson}{Turnbull and
  Watson}{1993}]%
        {turnbull1993maps}
\bibfield{author}{\bibinfo{person}{David Turnbull} {and} \bibinfo{person}{Helen
  Watson}.} \bibinfo{year}{1993}\natexlab{}.
\newblock \bibinfo{booktitle}{\emph{Maps Are Territories Science is an Atlas: A
  Portfolio of Exhibits}}.
\newblock \bibinfo{publisher}{University of Chicago Press}.
\newblock


\bibitem[\protect\citeauthoryear{Vattam, Wiltgen, Helms, Goel, and Yen}{Vattam
  et~al\mbox{.}}{2011}]%
        {vattam2011dane}
\bibfield{author}{\bibinfo{person}{Swaroop Vattam}, \bibinfo{person}{Bryan
  Wiltgen}, \bibinfo{person}{Michael Helms}, \bibinfo{person}{Ashok~K Goel},
  {and} \bibinfo{person}{Jeannette Yen}.} \bibinfo{year}{2011}\natexlab{}.
\newblock \showarticletitle{{DANE: fostering creativity in and through
  biologically inspired design}}.
\newblock In \bibinfo{booktitle}{\emph{Design Creativity 2010}}.
  \bibinfo{publisher}{Springer}, \bibinfo{pages}{115--122}.
\newblock


\bibitem[\protect\citeauthoryear{Verbert, Govaerts, Duval, Santos, Assche,
  Parra, and Klerkx}{Verbert et~al\mbox{.}}{2014}]%
        {verbert2014learning}
\bibfield{author}{\bibinfo{person}{Katrien Verbert}, \bibinfo{person}{Sten
  Govaerts}, \bibinfo{person}{Erik Duval}, \bibinfo{person}{Jose~Luis Santos},
  \bibinfo{person}{Frans Assche}, \bibinfo{person}{Gonzalo Parra}, {and}
  \bibinfo{person}{Joris Klerkx}.} \bibinfo{year}{2014}\natexlab{}.
\newblock \showarticletitle{{Learning dashboards: an overview and future
  research opportunities}}.
\newblock \bibinfo{journal}{\emph{Personal and Ubiquitous Computing}}
  \bibinfo{volume}{18}, \bibinfo{number}{6} (\bibinfo{year}{2014}),
  \bibinfo{pages}{1499--1514}.
\newblock


\bibitem[\protect\citeauthoryear{Vyas, van~der Veer, and Nijholt}{Vyas
  et~al\mbox{.}}{2013}]%
        {vyas2013creative}
\bibfield{author}{\bibinfo{person}{Dhaval Vyas}, \bibinfo{person}{Gerrit
  van~der Veer}, {and} \bibinfo{person}{Anton Nijholt}.}
  \bibinfo{year}{2013}\natexlab{}.
\newblock \showarticletitle{{Creative practices in the design studio culture:
  collaboration and communication}}.
\newblock \bibinfo{journal}{\emph{Cognition, Technology {\&} Work}}
  \bibinfo{volume}{15}, \bibinfo{number}{4} (\bibinfo{year}{2013}),
  \bibinfo{pages}{415--443}.
\newblock


\bibitem[\protect\citeauthoryear{Wang, Yang, Jin, Shechtman, Agarwala, Brandt,
  and Huang}{Wang et~al\mbox{.}}{2015}]%
        {wang2015deepfont}
\bibfield{author}{\bibinfo{person}{Zhangyang Wang}, \bibinfo{person}{Jianchao
  Yang}, \bibinfo{person}{Hailin Jin}, \bibinfo{person}{Eli Shechtman},
  \bibinfo{person}{Aseem Agarwala}, \bibinfo{person}{Jonathan Brandt}, {and}
  \bibinfo{person}{Thomas~S Huang}.} \bibinfo{year}{2015}\natexlab{}.
\newblock \showarticletitle{{Deepfont: Identify your font from an image}}. In
  \bibinfo{booktitle}{\emph{Proceedings of the 23rd ACM international
  conference on Multimedia}}. ACM, \bibinfo{pages}{451--459}.
\newblock


\bibitem[\protect\citeauthoryear{Webb}{Webb}{1982}]%
        {webb1982student}
\bibfield{author}{\bibinfo{person}{Noreen~M Webb}.}
  \bibinfo{year}{1982}\natexlab{}.
\newblock \showarticletitle{{Student interaction and learning in small
  groups}}.
\newblock \bibinfo{journal}{\emph{Review of Educational research}}
  \bibinfo{volume}{52}, \bibinfo{number}{3} (\bibinfo{year}{1982}),
  \bibinfo{pages}{421--445}.
\newblock


\bibitem[\protect\citeauthoryear{Whyte, Ewenstein, Hales, and Tidd}{Whyte
  et~al\mbox{.}}{2007}]%
        {whyte2007visual}
\bibfield{author}{\bibinfo{person}{Jennifer~K Whyte}, \bibinfo{person}{Boris
  Ewenstein}, \bibinfo{person}{Michael Hales}, {and} \bibinfo{person}{Joe
  Tidd}.} \bibinfo{year}{2007}\natexlab{}.
\newblock \showarticletitle{{Visual practices and the objects used in design}}.
\newblock \bibinfo{journal}{\emph{Building Research {\&} Information}}
  \bibinfo{volume}{35}, \bibinfo{number}{1} (\bibinfo{year}{2007}),
  \bibinfo{pages}{18--27}.
\newblock


\bibitem[\protect\citeauthoryear{Windisch}{Windisch}{1941}]%
        {windisch1941schule}
\bibfield{author}{\bibinfo{person}{Hans Windisch}.}
  \bibinfo{year}{1941}\natexlab{}.
\newblock \bibinfo{booktitle}{\emph{{Schule der Farbenphotographie}}}.
\newblock \bibinfo{publisher}{Heering-Verlag}.
\newblock


\bibitem[\protect\citeauthoryear{Woodruff}{Woodruff}{2019}]%
        {woodruff201910things}
\bibfield{author}{\bibinfo{person}{Allison Woodruff}.}
  \bibinfo{year}{2019}\natexlab{}.
\newblock \showarticletitle{10 things you should know about algorithmic
  fairness}.
\newblock \bibinfo{journal}{\emph{interactions}} \bibinfo{volume}{26},
  \bibinfo{number}{4} (\bibinfo{year}{2019}), \bibinfo{pages}{47--51}.
\newblock


\bibitem[\protect\citeauthoryear{Woodruff, Fox, Rousso-Schindler, and
  Warshaw}{Woodruff et~al\mbox{.}}{2018}]%
        {woodruff2018qualitative}
\bibfield{author}{\bibinfo{person}{Allison Woodruff}, \bibinfo{person}{Sarah~E
  Fox}, \bibinfo{person}{Steven Rousso-Schindler}, {and}
  \bibinfo{person}{Jeffrey Warshaw}.} \bibinfo{year}{2018}\natexlab{}.
\newblock \showarticletitle{A qualitative exploration of perceptions of
  algorithmic fairness}. In \bibinfo{booktitle}{\emph{Proceedings of the 2018
  CHI Conference on Human Factors in Computing Systems}}.
  \bibinfo{pages}{1--14}.
\newblock


\bibitem[\protect\citeauthoryear{Wrigley and Straker}{Wrigley and
  Straker}{2017}]%
        {wrigley2017design}
\bibfield{author}{\bibinfo{person}{Cara Wrigley} {and} \bibinfo{person}{Kara
  Straker}.} \bibinfo{year}{2017}\natexlab{}.
\newblock \showarticletitle{{Design thinking pedagogy: The educational design
  ladder}}.
\newblock \bibinfo{journal}{\emph{Innovations in Education and Teaching
  International}} \bibinfo{volume}{54}, \bibinfo{number}{4}
  (\bibinfo{year}{2017}), \bibinfo{pages}{374--385}.
\newblock


\bibitem[\protect\citeauthoryear{Yal{\c{c}}{\i}n, Elmqvist, and
  Bederson}{Yal{\c{c}}{\i}n et~al\mbox{.}}{2017}]%
        {yalccin2017keshif}
\bibfield{author}{\bibinfo{person}{Mehmet~Adil Yal{\c{c}}{\i}n},
  \bibinfo{person}{Niklas Elmqvist}, {and} \bibinfo{person}{Benjamin~B
  Bederson}.} \bibinfo{year}{2017}\natexlab{}.
\newblock \showarticletitle{Keshif: Rapid and expressive tabular data
  exploration for novices}.
\newblock \bibinfo{journal}{\emph{IEEE transactions on visualization and
  computer graphics}} \bibinfo{volume}{24}, \bibinfo{number}{8}
  (\bibinfo{year}{2017}), \bibinfo{pages}{2339--2352}.
\newblock


\bibitem[\protect\citeauthoryear{Yang, Steinfeld, Ros{\'e}, and Zimmerman}{Yang
  et~al\mbox{.}}{2020}]%
        {yang2020re}
\bibfield{author}{\bibinfo{person}{Qian Yang}, \bibinfo{person}{Aaron
  Steinfeld}, \bibinfo{person}{Carolyn Ros{\'e}}, {and} \bibinfo{person}{John
  Zimmerman}.} \bibinfo{year}{2020}\natexlab{}.
\newblock \showarticletitle{Re-examining Whether, Why, and How Human-AI
  Interaction Is Uniquely Difficult to Design}. In
  \bibinfo{booktitle}{\emph{Proceedings of the 2020 chi conference on human
  factors in computing systems}}. \bibinfo{pages}{1--13}.
\newblock


\bibitem[\protect\citeauthoryear{Yang, Steinfeld, and Zimmerman}{Yang
  et~al\mbox{.}}{2019}]%
        {yang2019unremarkable}
\bibfield{author}{\bibinfo{person}{Qian Yang}, \bibinfo{person}{Aaron
  Steinfeld}, {and} \bibinfo{person}{John Zimmerman}.}
  \bibinfo{year}{2019}\natexlab{}.
\newblock \showarticletitle{Unremarkable ai: Fitting intelligent decision
  support into critical, clinical decision-making processes}. In
  \bibinfo{booktitle}{\emph{Proceedings of the 2019 CHI Conference on Human
  Factors in Computing Systems}}. \bibinfo{pages}{1--11}.
\newblock


\bibitem[\protect\citeauthoryear{Yilmaz and Daly}{Yilmaz and Daly}{2016}]%
        {yilmaz2016feedback}
\bibfield{author}{\bibinfo{person}{Seda Yilmaz} {and} \bibinfo{person}{Shanna~R
  Daly}.} \bibinfo{year}{2016}\natexlab{}.
\newblock \showarticletitle{Feedback in concept development: Comparing design
  disciplines}.
\newblock \bibinfo{journal}{\emph{Design Studies}}  \bibinfo{volume}{45}
  (\bibinfo{year}{2016}), \bibinfo{pages}{137--158}.
\newblock


\bibitem[\protect\citeauthoryear{Young, Hazarika, Poria, and Cambria}{Young
  et~al\mbox{.}}{2018}]%
        {young2018recent}
\bibfield{author}{\bibinfo{person}{Tom Young}, \bibinfo{person}{Devamanyu
  Hazarika}, \bibinfo{person}{Soujanya Poria}, {and} \bibinfo{person}{Erik
  Cambria}.} \bibinfo{year}{2018}\natexlab{}.
\newblock \showarticletitle{Recent trends in deep learning based natural
  language processing}.
\newblock \bibinfo{journal}{\emph{ieee Computational intelligenCe magazine}}
  \bibinfo{volume}{13}, \bibinfo{number}{3} (\bibinfo{year}{2018}),
  \bibinfo{pages}{55--75}.
\newblock


\bibitem[\protect\citeauthoryear{Yu, Yang, Liu, Song, Xiang, and Hospedales}{Yu
  et~al\mbox{.}}{2017}]%
        {yu2017sketch}
\bibfield{author}{\bibinfo{person}{Qian Yu}, \bibinfo{person}{Yongxin Yang},
  \bibinfo{person}{Feng Liu}, \bibinfo{person}{Yi-Zhe Song},
  \bibinfo{person}{Tao Xiang}, {and} \bibinfo{person}{Timothy~M Hospedales}.}
  \bibinfo{year}{2017}\natexlab{}.
\newblock \showarticletitle{Sketch-a-net: A deep neural network that beats
  humans}.
\newblock \bibinfo{journal}{\emph{International journal of computer vision}}
  \bibinfo{volume}{122}, \bibinfo{number}{3} (\bibinfo{year}{2017}),
  \bibinfo{pages}{411--425}.
\newblock


\bibitem[\protect\citeauthoryear{Yuan, Luther, Krause, Vennix, Dow, and
  Hartmann}{Yuan et~al\mbox{.}}{2016}]%
        {yuan2016almost}
\bibfield{author}{\bibinfo{person}{Alvin Yuan}, \bibinfo{person}{Kurt Luther},
  \bibinfo{person}{Markus Krause}, \bibinfo{person}{Sophie~Isabel Vennix},
  \bibinfo{person}{Steven~P Dow}, {and} \bibinfo{person}{Bjorn Hartmann}.}
  \bibinfo{year}{2016}\natexlab{}.
\newblock \showarticletitle{{Almost an expert: The effects of rubrics and
  expertise on perceived value of crowdsourced design critiques}}. In
  \bibinfo{booktitle}{\emph{Proceedings of the 19th ACM Conference on
  Computer-Supported Cooperative Work {\&} Social Computing}}. ACM,
  \bibinfo{pages}{1005--1017}.
\newblock


\bibitem[\protect\citeauthoryear{Zimmerman, Forlizzi, and Evenson}{Zimmerman
  et~al\mbox{.}}{2007}]%
        {zimmerman2007research}
\bibfield{author}{\bibinfo{person}{John Zimmerman}, \bibinfo{person}{Jodi
  Forlizzi}, {and} \bibinfo{person}{Shelley Evenson}.}
  \bibinfo{year}{2007}\natexlab{}.
\newblock \showarticletitle{{Research through design as a method for
  interaction design research in HCI}}. In
  \bibinfo{booktitle}{\emph{Proceedings of the ACM CHI}}. ACM,
  \bibinfo{pages}{493--502}.
\newblock


\end{thebibliography}

\appendix
\small

\section{Discussion Topics and Questions}
\label{sec:discussion_topics}
We engaged design instructors in an ongoing discourse\textemdash spanning a period of more than one year\textemdash discussing their course learning objective, assignment specifications, assessment and feedback practices.
We discussed whether and how computational means could support them in their assessment and feedback.
We term these as discussions because we shared perspectives that we gained from prior work and by engaging with other instructors.
In addition to below, emergent topics included students’ use of sketching, instructors’ creativity assessment, and students’ considering instructor feedback beyond a specific deliverable.

\subsection{Individual Discussion}

\begin{itemize}
    
\item What design courses do you teach?

\item How do you define learning objectives for each course? What are the guiding principles?

\item As a design instructor, why do you think these specific objectives are important?

\item How do you prioritize?

\item What assignments do you define for meeting the objectives?

\item Do the assignments vary at times? How? Why?

\item How do you formulate the grading rubric for an assignment?

\item Are there weights for the design product and process?

\item What is given more weight in assignments: understanding course material, application of learned principles, visual design, creative ideas, other factors?

\item Do weights vary across assignments? How?

\item How are grading rubrics applied? Are there macro and micro level guidelines based on the learning objectives of that assignment?

\item What instructions are TAs / graders given toward applying the rubric? How is consistency ensured?

\item Do you use peer review for grading? Do the guidelines differ? How?

\item How do you understand the picture of a class? for e.g., How is a class doing overall? What are common and specific problems?

\item How do you address the common and specific problems?

\item Do you think that the design process / addressing of problems could be better supported by seeing students' progress more regularly?

\item What do you think about our intention to design computer algorithms that measure aspects of design work that are representative of the criteria in your grading rubrics?

\item Would some kind of analytics based on the grading rubric be helpful?

\item Would some representations of their work be useful, e.g. conceptual themes, assignment sections?

\item Would a dashboard interface help in this regard (show a mockup)?

\item What else could be useful?

\item What features would you like to see? Examples, sorting, mean, median, outliers?

\item What other features do you think would be useful?

\end{itemize}

\subsection{Workshop Discussion}

For workshop, while we had put together a list of topics, 
the discussion emerged on its own after instructors finished presenting their design ideation assignments, along with assessment approach. Topics that emerged included:

\begin{itemize}
     
\item the sizes of courses, teaching teams, and student teams;

\item the level of details in assignment and rubric specifications;

\item challenges in tracking students’ progress and incorporation of feedback; and

\item difficulty assessing contributions in team assignments.

\end{itemize}

\section{Design Characteristics Across Fields}
\label{sec:design_characteristics}
In Section \ref{sec:design_creativity_analytics}, we developed a set of situated design creativity analytics\textemdash Fluency, Flexibility, Visual Consistency, Multiscale Organization, and Legible Contrast\textemdash that can provide AI-based, on demand, actionable assessment to design students.
Here, we enumerate additional characteristics identified through co-design engagements with design instructors across fields. 

{
\renewcommand{\arraystretch}{1.3} 
\begin{table*}[htb]
\center
\footnotesize
\begin{tabular}{c|c|c|c|c|c}
{\bf Analytic} & {\bf \makecell{Landscape \\ Architecture \& \\ Urban Planning}} & {\bf \makecell{Computer \\ Science and \\ Engineering}} & {\bf \makecell{Mechanical \\ Engineering}} & {\bf \makecell{Interactive \\ Art \& Design}} & {\bf Architecture} \\
\hline
Fluency & & & \myBullet{} & \myBullet{} &  \\ \rowcolor[gray]{0.95}
Flexibility & & \myBullet{} & \myBullet{} & \myBullet{} &  \\
Visual Consistency & \myBullet{} & \myBullet{} & & & \myBullet{} \\ \rowcolor[gray]{0.95}
Multiscale Organization & \myBullet{} & \myBullet{} & & &  \\
Legible Contrast & \myBullet{} & \myBullet{} & & &  \\ \rowcolor[gray]{0.95}
Team Members' Contribution & \myBullet{} & \myBullet{} & \myBullet{} & \myBullet{} &  \\
\arrayrulecolor{gray}\hline
Novelty & & \myBullet{} & \myBullet{} & &  \\ \rowcolor[gray]{0.95}
Relative Object Sizes & \myBullet{} & & & & \myBullet{} \\
Number of Iterations & \myBullet{} & & \myBullet{} & &  \\ \rowcolor[gray]{0.95}
Use of Small Multiples \cite{tufte1990envisioning} & \myBullet{} & & & &  \\
Range of Colors & \myBullet{} & & & &  \\ \rowcolor[gray]{0.95}
Coverage of Assigned Sources & & \myBullet{} & & &  \\
Edits during Practice & \myBullet{} & & & &  \\ \rowcolor[gray]{0.95}
\end{tabular}
\caption{Learning analytics relevant to design education. In the top part, we list analytics for which we developed AI-based approaches in the paper. In the bottom part, we list additional characteristics that instructors in our study emphasize. We prioritized developing analytics based on: a) the number of contexts which emphasize a design characteristic and b) the extent of conceptual and empirical evidence we identified in regard to developing an effective AI approach for the corresponding analytic.
}
\label{tab:potential_analytics}
\end{table*}
}

\end{document}